\documentclass[11pt,a4paper]{article}
\usepackage{jcappub}
\pdfoutput=1 
\usepackage[T1]{fontenc}
\usepackage{rviewport}
\usepackage{graphicx}	
\usepackage{amsmath}	
\usepackage{amssymb}	
\usepackage{xspace}
\usepackage{mathtools}
\usepackage{url}
 
\newcommand{\vareps}{\varepsilon}
\newcommand{\Fermi}{\emph{Fermi}\xspace}
\newcommand{\Swift}{\emph{Swift}\xspace}
\newcommand{\jetset}{{\scshape{Jetset}}\xspace}
\newcommand{\TXS}{TXS\,0506+056\xspace} 
\newcommand{\PKS}{PKS\,1502+106\xspace}
\newcommand{\icnu}{IC-190730A\xspace}
\newcommand{\icnuold}{IC-170922A\xspace}

\newcommand{\gRay}{$\gamma$-ray\xspace}
\newcommand{\gRays}{$\gamma$-rays\xspace}

\newcommand{\blr}{\rm BLR}
\newcommand{\ergs}{$\rm~erg~s^{-1}$\xspace}
\newcommand{\ergscmsq}{$\rm~erg~s^{-1}~cm^{-2}$\xspace}
\newcommand{\chisq}{$\chi^2$\xspace}

\newcommand{\be}{\begin{equation}}
\newcommand{\ee}{\end{equation}}
\newcommand{\ba}{\begin{eqnarray}}
\newcommand{\ea}{\end{eqnarray}}
\newcommand{\bi}{\begin{itemize}}
\newcommand{\ei}{\end{itemize}}

\title{Multi-messenger emission from the parsec-scale jet of the flat-spectrum radio quasar \PKS coincident with high-energy neutrino IceCube-190730A}

\author[a,1]{Foteini Oikonomou,\note{Corresponding author.}}
\author[b]{Maria Petropoulou,}
\author[c,d,e,f]{Kohta Murase,}
\author[g]{Aaron Tohuvavohu,}
\author[h,i]{Georgios Vasilopoulos,}
\author[j]{Sara Buson,}
\author[k]{and Marcos Santander}

\affiliation[a]{Institutt for Fysikk, Norwegian University of Science and Technology, Trondheim, Norway}

\affiliation[b]{Department of  Physics, National and Kapodistrian University of Athens, 
Greece}
\affiliation[c]{Department of Physics, Pennsylvania State University,
University Park, USA}
\affiliation[d]{Department of Astronomy \& Astrophysics,
Pennsylvania State University,
University Park, USA}
\affiliation[e]{Center for Multimessenger Astrophysics,
Institute for Gravitation and the Cosmos,
Pennsylvania State University,
University~Park, USA}
\affiliation[f]{Center for Gravitational Physics, Yukawa Institute for
Theoretical Physics, Kyoto, Japan}
\affiliation[g]{Department of Astronomy \& Astrophysics, University of Toronto, Toronto, Canada}
\affiliation[h]{Department of Astronomy, Yale University, New Haven, USA}
\affiliation[i]{Université de Strasbourg, CNRS, Observatoire astronomique de Strasbourg, Strasbourg, 
France}
\affiliation[j]{Institut f\"{u}r Theoretische Physik und Astrophysik, 
Julius-Maximilian Universit\"{a}t W\"{u}rzburg, \\ Germany} 
\affiliation[k]{Department of Physics and Astronomy,
University of Alabama,
Tuscaloosa, USA}
\date{\today}

\emailAdd{foteini.oikonomou@ntnu.no}
\emailAdd{mpetropo@phys.uoa.gr}
\emailAdd{murase@psu.edu}
\emailAdd{tohuvavohu@astro.utoronto.ca}
\emailAdd{georgios.vasilopoulos@astro.unistra.fr}
\emailAdd{sara.buson@uni-wuerzburg.de}
\emailAdd{jmsantander@ua.edu}

\abstract
{On July 30th, 2019 IceCube detected a high-energy astrophysical muon neutrino candidate, \icnu with a $67\%$ probability of astrophysical origin. The flat spectrum radio quasar (FSRQ) PKS 1502\,+106 is in the error circle of the neutrino. Motivated by this observation, we study \PKS as a possible source of \icnu. \PKS was in a quiet state in terms of UV/optical/X-ray/\gRay flux at the time of the neutrino alert, we therefore model the expected neutrino emission from the source during its average long-term state, and investigate whether the emission of \icnu as a result of the quiet long-term emission of \PKS is plausible. We analyse UV/optical and X-ray data and collect additional observations from the literature to construct the multi-wavelength spectral energy distribution of \PKS. We perform leptohadronic modelling of the multi-wavelength emission of the source and determine the most plausible emission scenarios and the maximum expected accompanying neutrino flux. A model in which the multi-wavelength emission of \PKS originates beyond the broad-line region and inside the dust torus is most consistent with the observations. In this scenario, \PKS can have produced up to of order one muon neutrino with energy exceeding 100 TeV in the lifetime of IceCube. An appealing feature of this model is that the required proton luminosity is consistent with the average required proton luminosity if blazars power the observed ultra-high-energy-cosmic-ray flux and well below the source's Eddington luminosity. If such a model is ubiquitous among FSRQs, additional neutrinos can be expected from other bright sources with energy $\gtrsim 10$~PeV.}


\notoc
\begin{document} 

\maketitle
\flushbottom
\section{Introduction}
\label{sec:intro} 

The IceCube Neutrino Observatory~\footnote{\url{http://icecube.wisc.edu}} reported the observation of a flux of neutrinos of 
astrophysical origin in 2013~\cite{PhysRevLett.111.021103,icecubeScience,PhysRevLett.113.101101}. Updated analyses with higher statistics have strengthened the significance of the observation~\cite{2019ICRC...36.1017S,Schneider19,Aartsen:2020aqd}. The origin of the neutrinos is an open question, though a plethora of models have been considered (see
e.g.~\cite{Ahlers:2018fkn} for a review). 

In 2017, the IceCube Collaboration reported the observation of a
$\gtrsim 290$~TeV muon neutrino, IceCube-170922A, coincident with a 6-month-long 
$\gamma$-ray flare of the blazar TXS\,0506+056~\cite{IceCube:2018dnn}  
at redshift $z = 0.3365$~\cite{Paiano:2018qeq}. The association of the neutrino with the flare of TXS 0506+056 is
inconsistent with arising by chance at the 3$\sigma$ level, making \TXS the first extragalactic astrophysical source to have been identified as a possible high-energy neutrino source at such a level of significance. 

Blazars are active galactic nuclei (AGN) with a relativistic
jet, oriented at a small angle with respect to the line of
sight~\cite{Urry:1995mg}. Blazars are traditionally
divided into two main sub-classes, namely BL~Lacertae objects (BL~Lacs) and Flat Spectrum Radio Quasars (FSRQs), based on the characteristics of their optical spectra. FSRQs are on average more luminous sources, and display broad,
strong emission lines, which reveal the presence of an efficient accretion disc~\cite{1976MNRAS.175..613S} and a so-called ``broad-line region'' of gas clouds, which intercept a fraction of the radiation from the accretion disk. BL~Lacs exhibit at most weak emission
lines, or in many cases featureless optical spectra, and the majority of them have a much less radiatively efficient accretion disk.
Blazars have long been discussed as some of the most likely sources of
high-energy neutrinos and cosmic rays, in parts due to their large luminosities, and due to the fact that they possess large, powerful jets, which are considered an ideal environment for particle acceleration (see~\cite{
1989A&A...221..211M,
1992A&A...260L...1M,
1993A&A...269...67M,
1995APh.....3..295M,
1997ApJ...488..669H,
Atoyan:2001ey,
2003APh....18..593M,
Murase:2011cy,
Dermer:2012rg,
2014PhRvD..90b3007M,
Padovani:2014bha,
Dermer:2014vaa,
Petropoulou:2015upa,
Padovani:2016wwn,Gao:2016uld,Rodrigues:2017fmu},
and references therein).

On July 30th, 2019, IceCube detected a high-energy astrophysical neutrino with the \verb|ICECUBE_Astrotrack_Gold| alert stream. The threshold astrophysical neutrino purity for such Gold alerts is $50\%$. The particular alert, \icnu, has $67\%$ ``signalness''~\cite{pks_atel,Aartsen_2017}, meaning that the neutrino is astrophysical in origin with $67\%$ probability. The most probable energy of the neutrino is $\sim$300 TeV~\cite{gcn_ic2}, assuming an $\vareps_{\nu}^{-2.19}$ neutrino spectrum. 

A search for interesting sources within the uncertainty region of the arrival direction of the neutrino revealed that the 
blazar \PKS lies within the $50\%$ uncertainty region, with an offset of 0.31 degrees from the best-fit neutrino location. \PKS is a very bright 
FSRQ
at redshift $z = 1.8385$~\cite{1977ApJ...215..427S,1980ApJ...235..361R}. The source is in the 4FGL, the Fourth \Fermi Source Catalog (4FGL J1504.4+1029) and is one of the $\sim 50$ brightest sources in terms of their flux therein~\cite{2020ApJS..247...33A}. It is also part of the Third Catalog of Hard \Fermi-LAT sources~\cite{TheFermi-LAT:2017pvy} (3FHL J1504.3+1030). \PKS belongs to the IceCube and ANTARES \textit{a priori} defined monitored source list~\cite{Aartsen:2018ywr}, which consists of 34 \gRay bright sources. The source exhibited a strong outburst in the \Fermi band in August 2008~\cite{Ciprini:2009kr}, and was the second brightest extragalactic \gRay source in the sky during this time. In 2009, renewed strong \gRay activity was observed, which lasted until the beginning of 2010. 

The positional association of \PKS with \icnu{} prompts us to investigate the theoretical expectation for neutrino emission from this very powerful FSRQ. The intrinsically high luminosity of FSRQs, which makes them the most-luminous persistent sources in the Universe, and the existence of photon fields related to the accretion disk makes FSRQs excellent candidate sources for neutrino production~\cite{Atoyan:2001ey,Dermer:2014vaa,2014PhRvD..90b3007M,Reimer:2015qdr,Rodrigues:2017fmu}. The actual neutrino output of any FSRQ, depends on several factors which are uncertain, namely the proton content of the jet, the maximum energy to which protons can be accelerated, and the location of the bulk of the jet's non-thermal emission with respect to the powerful photon fields associated with the accretion disk. If the blazar emitting region is close to the base of the jet and significant amounts of protons are present, then the photon fields associated with the accretion disk can act as target fields for photopion interactions, allowing for the production of large neutrino fluxes. The location of the emitting region is an open question and may vary from source to source, but detailed astronomical observations can often provide interesting constraints as we detail in later sections. 

In what follows, we examine neutrino production in \PKS in a comprehensive set of scenarios, taking the observational constraints for the location of the high-energy emission region of the source into account. At the time of detection of \icnu, \PKS was in a quiet state in terms of its optical, UV, X-ray and \gRay flux, as will be detailed in section~\ref{sec:obs}.  Motivated by this fact, in this work we model the long-term quiet neutrino emission of \PKS and investigate whether \icnu could have been produced by the source during its long-term (quiescent activity). This should be contrasted to the case of  IC-170922A, which was detected while \TXS was undergoing its largest \gRay flare and exhibiting flaring activity in optical and X-rays. In this regard,the scenarios that we will investigate here are different from those applied to the detection of \icnuold and associated emission from \TXS in 2017.
 
Similarly, in 2014-15, \TXS must have exhibited major flaring activity in the X-ray to MeV \gRay energy range in order for the model predictions to get close to the observed neutrino flux~\cite{Reimer:2018vvw,Rodrigues:2018tku,Petropoulou:2019zqp,Zhang:2019htg}, though such a high-state cannot be confirmed by existing observations (due to, for example, lack of spectral coverage in the MeV energy range). As we will demonstrate in this paper, \PKS produces sufficient neutrino flux as to account for the observation of \icnu during its long-term ``quiet'' emission in  several of the scenarios we investigate (see also~\citealp{Rodrigues:2020fbu}). 

\PKS was in a long-term radio outburst which started in 2014 and reached an all-time high at the time of arrival of \icnu in terms of its 15~GHz flux; a similar behaviour was observed in \TXS at the time of arrival of \icnuold in 2017. In what follows, we review existing observations that constrain the  distance of the multiwavelength emission region  in the jet of \PKS. According to those the \gRay emission is very unlikely to be produced in the same region as the 15 GHz emission. In this work, we focus on scenarios in which the neutrinos are cospatially produced with the optical to \gRay emission, in a compact region, relatively close to the base of the jet. 

In section~\ref{sec:obs} we give details of the multiwavelength data used in this study. In section~\ref{sec:location} we summarise existing constraints on the location of the high-energy emitting region of \PKS, which, as we will see, has important implications for the expected neutrino emission. In section~\ref{sec:model} we give details of the theoretical framework used to model the neutrino emission from \PKS in this work. In section~\ref{sec:results} we present our results, and we summarise our findings in section~\ref{sec:discussion}. 

We assume a flat Universe with $H_0 = 70~\rm km~\rm s^{-1}~ Mpc^{-1}$, $\Omega_{\rm M} = 0.7$ and $\Omega_{\rm \Lambda} = 0.3$, placing \PKS at luminosity distance $d_{\rm L} = 14013.5$~Mpc. In what follows, primed quantities refer to the jet comoving frame, and unprimed quantities correspond to the observer frame. 

\section{Multiwavelength Observations}
\label{sec:obs} 

\subsection{Observations of the source at the time of the IceCube Alert}

Following the IceCube alert, the \textit{Neil Gehrels Swift Observatory} observed the field of \icnu on two epochs: the first one beginning on July 30, 21:57:43 UT, about 1.1 hours after the neutrino detection, and a follow-up exposure starting on August 1, 00:49:52 UT~\cite{xrt_atel}. The 0.3 - 10 keV X-ray flux of \PKS measured with the \Swift-X-ray Telescope (XRT)~\cite{xrt2005} was found to be lower than the long-term average listed in the \Swift-XRT point source (1SXPS) catalogue~\cite{2013yCat.9043....0E}, while the X-ray photon index was found to be consistent with that reported in the 1SXPS. 

Additional follow-up UV and optical observations with the \Swift Ultra-Violet Optical Telescope (UVOT)~\cite{uvot2005}, the MASTER Global Robotic Net~\cite{master_atel}, the Zwicky Transient Facility~\cite{ztf_atel}, the Gravitational-wave Optical Transient Observer~\cite{2019GCN.25255....1S}, and the University of Alabama 0.4m telescope~\cite{alabama_atel}, confirmed that source was not experiencing a flux enhancement at these wavelengths. 

The flux density of \PKS at 15~GHz measured with the OVRO 40m Telescope shows a long-term outburst that started in 2014. At the time of the \icnu alert, it was reaching an all-time high since the beginning of the OVRO monitoring in 2008 (about 4~Jy)~\cite{ovro_atel}. A similar 15 GHz long-term outburst was seen in \TXS during the neutrino event \icnuold~\cite[e.g.][]{Hovatta:2020lor}. 

\subsection{Long-term Observations} 
\label{subsec:data}
To construct the long-term, non-flaring spectral energy distribution (SED) of \PKS we analysed data from the \Swift UVOT and XRT as detailed below. In addition, we used data obtained from the SSDC online interactive archive~\footnote{\url{http://www.asdc.asi.it}} and the \Fermi-LAT spectrum obtained by the analysis of~\cite{franckowiak2020patterns}. The \gRay emission of \PKS observed with the \Fermi-LAT at energy exceeding $\sim 15$~GeV is affected by interactions with the extragalactic background light (EBL). We correct for the effect of the EBL using the EBL model of ~\cite{Franceschini:2008tp}.

\begin{figure}
\includegraphics[width = \textwidth]{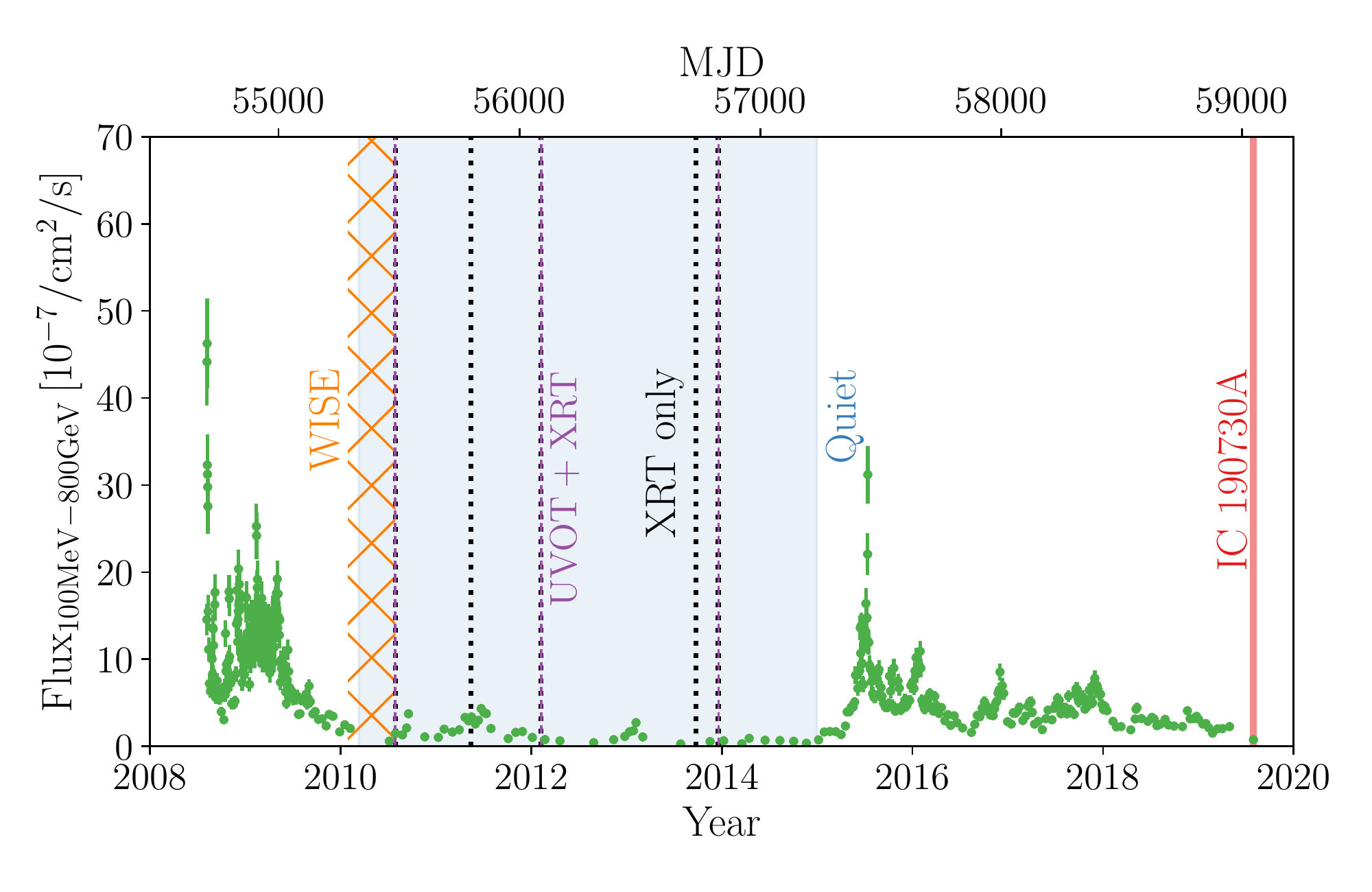}
\caption{The \Fermi-LAT light curve of \PKS in the 100 MeV-800 GeV energy range from the analysis of~\cite{franckowiak2020patterns}. The blue shaded band indicates the quiet period studied in this work. The red solid line denotes the time of detection of \icnu. The purple dashed lines denote times of XRT and UVOT observations that coincide with the studied period and black dotted lines denote times of XRT only observations. The yellow hatched lines denote the approximate time of the WISE observations.
\label{fig:lightcurve}}
\end{figure}

Since we are interested in the long-term quiescent emission of \PKS we constructed the SED with data from the epoch MJD 55266-57022. 
This corresponds to a long quiet period in \gRays between 2010-2014 that was analysed by~\cite{franckowiak2020patterns}. 
Figure~\ref{fig:lightcurve} shows the \Fermi-LAT light curve of \PKS obtained by the analysis of~\cite{franckowiak2020patterns}, and reveals that the \gRay flux of \PKS around the time 
of detection of \icnu (shown with a red vertical line) was at a similar level as during
 the epoch MJD 55266-57022. The same is true for the X-ray and \gRay fluxes, as can be seen in figure 10 of~\cite{franckowiak2020patterns}. 

Below we give details of the analysis of the \Swift XRT and UVOT archival data and our estimate of the chance probability of association of \icnu with \PKS and the neutrino flux implied by the IceCube neutrino alert event \icnu. 

\subsubsection{\Swift-UVOT}

The \Swift-UVOT data covers $>100$ individual exposures over $>12$ years of monitoring the source in the six UVOT lenticular filters (\textit{v,b,u,uvw1,uvm2,uvw2}). These data were analyzed using the standard \texttt{uvotsource} of HEASoft (v6.26). The \texttt{uvotsource} tool performs aperture photometry \cite{uvotcalib} using user-specified source and background regions. A 5-arcsecond radius aperture was used for the foreground (source) region and a 25-arcsecond radius background region was defined in a nearby location with no evidence for any sources. The data were calibrated using the latest UVOT \texttt{CALDB} files.
The mean magnitudes were computed for all the UVOT observations during the epoch MJD 55266-57022. Error bars give the standard error on the mean. The optical/UV magnitudes were corrected for Galactic extinction using an $E(B-V)$ value of 0.0275 from~\cite{2011ApJ...737..103S}. Extinction coefficients at the central wavelength of each UVOT filter were calculated following the extinction law of~\cite{1999PASP..111...63F} with $R_v = 3.07$ using the York Extinction Solver~\cite{2004AJ....128.2144M}. 

\subsubsection{\Swift-XRT}

\Swift-XRT data were retrieved and analyzed though the UK \Swift \ science data centre~\footnote{\url{http://www.swift.ac.uk/user_objects/}}, using standard procedures as outlined in \cite{2007A&A...469..379E,2009MNRAS.397.1177E}. Light curves were created using average count rates for individual observations \cite{2007A&A...469..379E}, while spectra were created by stacking events within the epoch of interest \cite{2009MNRAS.397.1177E}. For spectral analysis, we grouped events using a minimum count per channel of one and used  {\tt xspec} \cite[version 12.9.0, ][]{1996ASPC..101...17A}.
We fitted an absorbed power-law model to the data using c-statistics \cite{1979ApJ...228..939C}.
The X-ray absorption was modeled using the {\tt TBabs} code, with elemental abundances adopted by \cite{2000ApJ...542..914W}, and atomic cross sections from \cite{1996ApJ...465..487V}.
The photoelectric absorption was fixed to the value of the Galactic foreground absorption.
We used a fixed column density~\footnote{Weighted average value around 0.1$^o$ cone, see online tool: \url{https://heasarc.gsfc.nasa.gov/cgi-bin/Tools/w3nh/w3nh.pl}} of N$_{\rm H{\rm, GAL}}$ = 2.03$\times10^{20}$ cm$^{-2}$ \cite{2016A&A...594A.116H}.
For the epoch MJD 55266-57022 we found a photon index $\Gamma$ of $1.68\pm0.3$.  
The error bars represent $1\sigma$ uncertainties. The average flux of \PKS reported
in the 2nd XRT Point Source Catalogue, 2XSPS~\cite{2020ApJS..247...54E} is based on 74 observations of the source between 2007-2018. It is a factor of four higher
than the flux we obtained for the quiet epoch MJD 55266-57022. The 2SXPS average index of the source is consistent with the photon index we obtained above within uncertainties.

\subsubsection{WISE}

During part of the epoch MJD 55266-57022, the \emph{Wide-field Infrared Survey Explorer} (\emph{WISE})
mission~\cite{2010AJ....140.1868W}, mapped the sky at infrared wavelengths, with photometry centered at wavelengths of 3.4, 4.6, 12, and 22~$\mu m$. \emph{WISE} photometric fluxes of \PKS are available as part of the \emph{AllWISE} data release and were retrieved from the SSDC database. Consistent values are reported in the NASA/IPAC Extragalactic Database. The angular resolution of WISE ranges between $6.1''$ (at 3.4~$\mu m$) to $12.0''$ (at 22~$\mu m$), thus the photometric fluxes inferred for a particular source can in principle be subject to contamination from nearby sources. In the case of \PKS, the WISE image of the source is well resolved and isolated, so such contamination is not expected. Additionally, the fact that archival, non-simultaneous, data obtained as part of the SDSS~\cite{2017ApJS..233...25A} and 2MASS surveys~\cite{2006AJ....131.1163S}, which we retrieved from the SSDC database, nicely connect the WISE and quiet UVOT data, as shown in figure~\ref{fig:SSC_SED_low_eta}, confirm that the WISE data are a fair description of the blazar's quiet state. The archival SDSS and 2MASS data are not taken into account in the analysis that follows. 

The epoch that is spanned by the data analysed and that we use as representative of \PKS at the time of arrival of \icnu is shown as a blue band in figure~\ref{fig:lightcurve}.

\subsection{Neutrino Observations} 
\subsubsection{Chance probability of association} 

For a rough estimate of the chance probability of association of \icnu with a \gRay emitting blazar, we consider the chance probability of finding any source from the 3FHL catalogue close to the arrival direction of the neutrino. There are 142 FSRQs (873 blazars) in the 3FHL. The chance probability of detecting a 3FHL source (FSRQ) within $\Delta \theta = 0.3^{\deg}$ of a single alert is
\be
p_{\rm chance} \sim \frac{N(\Delta \theta_{\rm source} < \Delta \theta) }{N_{\rm tot}}   \sim \frac{\pi \Delta \theta^2}{4\pi} N_{\rm source} 
\ee

\be
p_{\rm chance} = 
\begin{cases} 0.0064  &  \rm all ~(N_{\rm source} = 873) \\ 
              0.0010  &  \rm FSRQ~(N_{\rm source} = 142). \\
\end{cases} 
\ee

\noindent In reality, IceCube muon neutrino alerts do not arrive from the entire sky. On the other hand, in our calculation, we include all the sources listed in the 3FHL whereas not all of them are visible to IceCube, so our approach is reasonable as long as the number of 3FHL sources in the IceCube EHE/Gold alert channel field of view does not differ much from the number expected for equal areas.

The cumulative binomial probability to see one or more associations from $N_{\rm alerts} = 10$, where $N_{\rm alerts}$ is the total number of Gold EHE alerts that were transmitted prior to and including \icnu is
\be
P (\rm 1~or~more) = \sum^{N_{\rm alert}}_{i = 1} \binom{N_{\rm alert}}{i} p_{\rm chance} ^i (1-p_{\rm chance})^{N_{\rm alert} - i}. \\ 
\ee
Therefore, the chance probability of a Gold/EHE neutrino alert in the direction of a 3FHL source is
\be
P (\rm 1~or~more) = \begin{cases} 0.062~(1.5 \sigma)~{\rm all} \\ 0.010~(2.3 \sigma)~{\rm FSRQ ~only}. \end{cases}
\ee
In other words, even if we limit ourselves to the FSRQs of the 3FHL catalogue the statistical significance of the association is not remarkable. 

On the other hand, considering the 3FHL energy flux of \PKS, which is $1.1384\times10^{-11}$~erg~cm$^{-2}$~s$^{-1}$, places it on the top $15\%$ strongest sources in terms of their flux in the catalogue. Previous to the \icnu{} alert, IceCube has not detected high-energy neutrinos in the direction of \PKS consistent with an astrophysical origin, even though the source is monitored as part of the {\it a priori} source list search program~\cite{Aartsen:2018ywr}. Using the best-fit parameters for the astrophysical neutrino flux, $\phi_{\rm astro}$ from~\cite{2019ICRC...36.1017S}, and the solid angle subtended by $r_{90}$ of \icnu we calculate the number of diffuse astrophysical neutrinos in the direction of \PKS as, $N_{\rm astro, diff}~=~\int_{300 \rm TeV}^{\infty}~\phi_{\rm astro}~A_{\rm eff}(\delta,E) \rm{d}\Omega \Delta T \sim 0.03( \Delta T/10~{\rm yr}) \ll 1$, where $A_{\rm eff}$ is the effective area of the IceCube GFU alert analysis. The fact that the expected diffuse astrophysical neutrino background in the direction of \PKS in ten years is well below one, and that \PKS is the only \Fermi-LAT detected source in the 4FGL-DR3 inside the error-circle of the neutrino, motivates our investigation on whether \icnu{} may have been emitted by \PKS during its long-term activity since the astrophysical background and thus the expected number of background/foreground sources is relatively low.

All in all, even though the statistical association is not highly significant, and is in any case, \emph{a posteriori}, \PKS is an interesting example of a typical powerful FSRQ, and it is worth investigating its capabilities as a possible source of \icnu. Additionally, \PKS is one of the first \emph{bonafide} FSRQs to have been associated with a high-energy muon neutrino with a high probability of astrophysical origin making the findings of this work relevant for future neutrino searches from FSRQs in general. 

\subsubsection{Neutrino flux estimate}
\label{subsec:neutrino_flux}
We first make a preliminary estimate of the mean flux implied by the observation of 1 muon neutrino event in IceCube in the direction of \PKS, using the \verb|EHE+GFU_Gold|~effective area published in~\cite{2019ICRC...36.1021B}. 

The number of signal-only, muon (and antimuon) neutrinos, $N_{\nu_{\mu}}$, detected by IceCube during a time-interval $\Delta \mathrm{T}$ at declination $\delta$ is,
\begin{equation}
N_{\nu_{\mu}} =
  \int_{\vareps_{\nu_{\mu},{\rm min}}}^{\vareps_{\nu_{\mu},{\rm max}}} {\rm d} \vareps_{\nu_{\mu}}A_{\rm eff}(\vareps_{\nu_{\mu}},\delta) \phi_{\nu_{\mu}}\Delta T,
\label{eq:Nnu}
\end{equation}
where $\vareps_{\nu,{\rm min}}$ and $\vareps_{\nu,{\rm max}}$ are the minimum and
maximum neutrino energy, and $\phi_{\nu_{\mu}}$, the muon neutrino flux, differential in energy. 

We find, considering the GFU effective area, that for a neutrino spectrum that follows a power-law $\vareps^{-\gamma}$ with index $\gamma = 2$, $90\%$ of detected neutrinos in the declination range $\delta = [-5^{\circ}, 30^{\circ}]$ observed with IceCube, in the \verb|GFU_All| ~ channel, would have energy in the range 20~TeV - 8~PeV. At the same time, only neutrinos above a certain probability of being astrophysical individually, which is proportional to energy, are transmitted as alert events. At present, the lowest energy neutrino to have been transmitted as a Gold neutrino alert had energy $\sim 80$~TeV.\footnote{\url{https://gcn.gsfc.nasa.gov/amon_icecube_gold_bronze_events.html}}. 

IceCube began operations on April 5th 2008 (MJD 54557). About 11.3 years had elapsed until the detection of \icnu (MJD 58694). During the first years the IceCube detector was partially deployed (see e.g.~\cite{Oikonomou:2019djc} table B1 for a summary of the timeline). For an estimate of the effective number of full-time equivalent years of IceCube live time, we multiply the effective area of IceCube in its final 86-string configuration $A_{\rm eff,IC86}$ by the ratio of the partial effective areas to $A_{\rm eff,IC86}$ at 300~TeV. We find that the total live time corresponds to approximately ten years of full-time operation.

Using eq.~(\ref{eq:Nnu}), $\vareps_{\nu_{\rm min}} = 80$ TeV, $\vareps_{\nu_{\rm max}} = 8$~PeV and $\Delta T = 10$~yr, 
we obtain an integrated, muon neutrino energy flux of $4 \times 10^{-12}$~\ergscmsq, corresponding to an average, integrated muon neutrino luminosity of ~$ 10^{47}~{\rm  erg~s^{-1}}$ based on the detection of one neutrino. We show the all flavour neutrino flux per logarithmic energy, in figure~\ref{fig:SSC_SED_low_eta}. We also show the Feldman Cousins all-flavour neutrino flux upper limit, at the $90\%$ confidence level (CL), for which we find $N_{\rm UL} (N_{\rm signal} = 1) = 4.36$ under the assumption of zero background. We only use the estimate presented in this section to indicate an approximate flux corresponding to the detection of one neutrino in the plots of the spectral energy distribution. Throughout the rest of this work we predict the expected number of neutrino events in IceCube given a specific model spectrum. 

\section{The location of the emitting region of \PKS}
\label{sec:location}

The location of the emitting region of \PKS is of importance for predicting the neutrino emission, as it informs us about the availability of external photon fields, which can act as targets for the production of neutrinos in photomeson interactions. In typical FSRQs, the strength and distance of 
external photon fields from the base of the jet, is, to a good degree, dictated by the power of the accretion disk. We, therefore, review existing observational constraints on FSRQs in general and on \PKS specifically in what follows. 

 Generally, FSRQs possess a luminous, radiatively efficient accretion disc~\cite{1973A&A....24..337S}. The FSRQ jet propagates through radiation fields produced by the gas clouds which reprocess the radiation from the disc, namely, the ultraviolet photons of the Broad Line Region (BLR) at a small distance from the jet base. At a larger distance from the jet base, a dust torus (DT) captures a fraction of the accretion disk radiation and re-emits it at infrared wavelengths. 

It is generally thought that the emitting region in the jet of FSRQs
is inside the BLR or at least inside the DT. In such a case, these 
radiation fields provide seed photons for inverse Compton 
scattering by relativistic electrons from the jet. This process is referred to as external-Compton (EC) emission, and is generally thought to power the high-energy emission observed from FSRQs, \cite{1994ApJ...421..153S,1992A&A...256L..27D,1996MNRAS.280...67G}. Until recently, it had been standard to assume that EC on BLR photons was the process powering the high-energy emission of the majority of FSRQs in leptonic scenarios~\cite{1998MNRAS.301..451G,2008MNRAS.387.1669G,Sikora:2009tp}. 

\PKS has a BLR with bolometric luminosity $L_{\rm BLR}\gtrsim 10^{45}$\ergs based on emission line measurements. \cite{2006ApJ...637..669L} estimated $L_{\rm BLR}\sim 3.7 \times 10^{45}$\ergs based on the MgII emission line profile, whereas \cite{Sbarrato:2011ps} estimated $L_{\rm BLR}\sim 2 \times 10^{45}$\ergs, and \cite{2012ApJ...748...49S} give  $L_{\rm BLR}\sim 10^{45}$\ergs. We conservatively adopt the latter, lower estimate in this work. From the measured $L_{\rm BRL}$ we can derive the accretion disk luminosity using the relation
\be
L_{\rm BLR} = f_{\rm BLR} L_{d},
\ee
where $f_{\rm BLR}$ is the fraction of the accretion disk luminosity, $L_{d}$ assumed to be intercepted and reprocessed by the BLR, typically $f_{\blr} \sim 0.1$. Thus for \PKS we obtain $L_{d} \sim 10^{46}$\ergs. The BLR radius of FSRQs is roughly $r_{\rm BLR} \sim 10^{17}~(L_{d}/10^{45}{\rm erg~s^{-1}})^{1/2}$~cm \cite{Ghisellini_2009}, which gives $r_{\rm BLR} \sim 3\times 10^{17}~{\rm cm}\sim 0.1$~pc for \PKS.  

Additionally, FSRQs possess a dust torus with luminosity 
\be
L_{\rm DT} = f_{\rm DT} L_{d},
\ee
where $f_{\rm DT}$ is the fraction of $L_{d}$ intercepted and re-emitted
by the torus at infrared infrared wavelengths. Typically, $f_{\rm DT} \sim 0.5$, which gives $L_{\rm DT} \approx 5 \times 10^{45}$\ergs for \PKS. 
The radius of
the infrared torus is, approximately~\cite{2000ApJ...545..107B,Sikora_2002} 
\be
r_{\rm DT} = 2.5 \times 10^{18}~(L_{d}/10^{45}{\rm erg~s^{-1}})^{1/2}~\rm cm, 
\ee
which corresponds to $r_{\rm DT} \sim 2.6$~pc for \PKS. 
\noindent The energy density of the IR torus in the frame comoving with the jet is thus
\be
U^{\prime}_{\rm DT} = \frac{f_{\rm DT}L_{d}\Gamma^2}{4 \pi r_{\rm DT}^2 c}.
\ee

\subsection{Information from the spectral energy distribution of \PKS}
 Clues as to the location of the emitting region of a blazar are offered by the characteristics of the SED. For example, a large Compton dominance is a typical characteristic of external Compton models, due to the large available radiation energy density in the frame comoving with the jet~\cite[e.g.][]{2008MNRAS.387.1669G}. However, as we will also demonstrate in subsequent sections in this work, the SED, generally, cannot unambiguously determine the location of the dissipation region.

Though in the past it was standard to assume that the dissipation region is close to the black hole and inside the BLR, detailed analysis of \Fermi-LAT data has provided evidence to the contrary. \cite{Costamante:2018anp} analysed the \gRay spectra of 106 \Fermi-detected blazars and found that the majority of the sources studied, must have \gRay emitting regions beyond the BLR as the \gRay spectra do not exhibit the characteristic spectral signature that should be induced by BLR absorption in their \gRay spectra. \cite{Meyer:2019kpy} did a similar analysis, focusing on \Fermi-LAT data of six bright blazars during flares, and reach the same conclusion; namely the \gRay spectra are inconsistent with emission from inside the BLR. Their analysis lead them to conclude that the \gRay emitting region is most likely at a distance $\sim 1$ pc from the central black hole.  

In what follows, we review existing constraints on the location of the dissipation region of \PKS. 

In the work of~\cite{Ciprini:2009kr} it was found that the high-energy peak of the SED during the August 2008 flare of \PKS is consistent with having been produced by a combination of synchrotron-self-Compton and external-Compton emission by an emitting region positioned outside (but not far from) the BLR. 

\cite{2017ApJS..228....1Z} modelled the SED of \PKS compiled with simultaneous and quasi-simultaneous observations obtained by~\cite{Giommi:2011sn} in July-August 2010 when \PKS had already entered a quiet \gRay epoch. For this epoch, these authors found a better fit to the SED of \PKS when the emitting region is beyond the BLR but inside the DT. A similar conclusion was reached by~\cite{Costamante:2018anp} whose sample included the long-term SED of \PKS. Very recently, another study that utilised an SED diagnostic and which included the long-term SED of \PKS among 62 studied sources, concluded that powerful FSRQ jets, including the jet of \PKS are consistent with dissipating their energy inside the dust torus, but inconsistent with dissipating the bulk of their emission inside the BLR~\cite{Harvey:2020wd}.

On the other hand, the analyses of~\cite{Ding_2019, acharyya2020locating,2021MNRAS.503.3145B} report rapid variability (timescale, $t_{\rm var,obs}\sim 1$h) during the 2009 and 2015 flares of \PKS as observed with \Fermi-LAT (in \cite{2021MNRAS.503.3145B} they obtain $\sim 1$~\rm day considering the long-term light curve of \PKS and reject smaller variability intervals due to being below the Nyquist limit). Such rapid variability constrains the emitting region to be very compact, with size $r_b \leq ct_{\rm var,obs}\delta(1+z) = 1.4 \times 10^{15}~{\rm cm}$, where $\delta$ is the Doppler factor of the jet. Assuming a conical jet with opening angle $\theta \sim 1/\delta$, and emitting region that covers the entire jet opening, the distance of the emitting region from the cental engine can be estimated as $r_{\rm diss} \sim r_b/\theta \sim ct_{\rm var,obs} \delta^2/(1+z) = 0.02~{\rm pc}$, and thus well within the BLR for typical values of the Doppler factor which is generally $\delta \leq 50$. 

However, it should be noted that there is evidence of emission well beyond the BLR, in 
a very compact emitting region, at least for one well studied FSRQ, namely 
4C+21.35~\cite{Tavecchio_11}, demonstrating that short-term variability alone is not in itself solid proof of the location of the emitting region at a small distance to the jet base. Furthermore,~\cite{acharyya2020locating} are finally inconclusive about the location of the emitting region of \PKS, as other parts of their analysis, for example, the shape of the \gRay spectrum, point to dissipation beyond the BLR.  

\subsection{Information from radio interferometric monitoring of \PKS and multifrequency correlations}

When available, very-long baseline interferometric radio observations of a blazar, which achieve excellent angular resolution, can determine the absolute distance of the detected radiation at a specific frequency from the base of the jet. This process does not suffer from the degeneracies inherent to the SED-based approach of determining the location of radiation dissipation. \PKS has been the subject of several very-long baseline interferometry campaigns which give detailed insight on the jet's properties and morphology (e.g.  \cite{An:2004cd,Ciprini:2009kr,Karamanavis_a,Karamanavis_b}). It is also one of the monitored sources of the F-GAMMA program~\cite{2016A&A...596A..45F}, which performed multifrequency radio monitoring of \gRay blazars. By identifying correlations between radio and \gRay light curves and fluxes, it was possible to determine the location of the \gRay production in the monitored jets. 

In the work of~\cite{Karamanavis_a} which analysed Very Long Baseline Interferometric (VLBI) and F-GAMMA observations of \PKS during the 2008-10 \gRay high state, an upper limit of $5.9$ pc was derived for the location of the \gRay emitting region from the SMBH. The subsequent work of~\cite{Karamanavis_b}, focused on the same period, and assuming the relation of~\cite{2014MNRAS.441.1899F} for the mean spatial separation of the radio-emitting core from the \gRay emitting region, these authors concluded that the \gRay emitting region of \PKS must be at a distance of $1.9\pm 1.1$~pc from the SMBH. This result places the \gRay emitting region of \PKS during the 2008-10 flare period beyond the BLR but inside the DT. With a similar approach,~\cite{10.1093/mnras/stu1749} calculated the time lag between the \gRay and 15 GHz radio emission. Combining their results with those of~\cite{Karamanavis_b} leads to inferred separation of $1.8\pm1.3$~pc of the \gRay active region from the jet base. 

In the analysis of~\cite{Shao:2020tth}, who searched for multi-frequency correlations in the long-term light curves of \PKS, it was concluded that the optical and \gRay emission of \PKS are produced co-spatially, within uncertainties. Together with the results of~\cite{Karamanavis_a} their work provides an estimate of the magnetic field strength in the optical and \gRay emitting region of \PKS, which was determined to be $\sim$0.36~G. 

In the recent work of~\cite{2021MNRAS.503.3145B} who analysed 15 GHz Very-Long Baseline Array (VLBA) data and astrometric 8 GHz VLBA data, it was found that the jet of \PKS exhibits unusual dynamics with some emission components appearing to have motion perpendicular to the jet axis. 

\section{Modelling of the SED of \PKS and predicted neutrino emission} 
\label{sec:model}

We consider scenarios for neutrino emission by \PKS, which differ mainly in terms 
of the location of the neutrino emitting region with respect to the base of the jet. 

Though the discussion on the location of the \gRay emitting region of \PKS in section~\ref{sec:location} is not fully conclusive, the balance of evidence 
points to the \gRay emitting region being beyond the BLR, likely inside the DT,
but is also consistent with being outside (beyond) the DT. 

We consider a one-zone synchrotron and synchrotron-self-Compton scenario, which means effectively that the emitting region is at a large distance from the jet base, beyond the BLR and DT, and one ``external Compton'' scenario, in which the emitting region is inside the DT but beyond the BLR. Based on the balance of observational evidence we consider it unlikely that the emitting region is inside the BLR, thus we do not investigate such a scenario. Such a case has nevertheless been treated in the work of~\cite{Rodrigues:2020fbu}. 

To keep the number of free parameters to a minimum, we assume that the neutrino emitting region is always cospatial with the emitting region of the radiation of \PKS, with the exception of the radio data, which must originate in the large scale jet of \PKS~(see~\cite{Karamanavis_a,Karamanavis_b}). As a reminder, astronomical observations of \PKS suggest co-spatial emission of \gRays and optical emission~\cite{Shao:2020tth}, supporting the ``one-zone'' picture for the non-flaring emission of this source in the optical to \gRay energy range.

In all the studied scenarios, neutrinos are produced by photomeson interactions of the protons with the photon fields of \PKS. We give details of the modelling approach in section~\ref{subsec:neutrino_model}, and details specific to each of the two models in Sects.~\ref{subsec:ssc_modelling} and~\ref{subsec:DT_modelling}, respectively. 

\subsection{Modelling approach}  
\label{subsec:neutrino_model} 

We model the long-term SED of \PKS under the assumption of a single, spherical emitting region in the jet which contains relativistic leptons and protons which are injected at a constant rate. To calculate the multimessenger emission of the source we use a multistep process. 

We model the emission of primary leptons using the \jetset code~\footnote{\url{https://jetset.readthedocs.io/en/latest/}} \cite{2006A&A...448..861M,2009A&A...501..879T,2011ApJ...739...66T}. \jetset models the interactions of relativistic leptons inside the blazar emitting region with internal (synchrotron) radiation and external photon fields and computes the escaping radiation spectrum. \jetset does not include the radiation emitted by photomeson interactions of hadrons in the blazar jet. Our treatment is valid, in the presence of relativistic hadrons, as long as the radiation emitted from the interactions of protons in the blazar jet, is subdominant to those of the leptons, which is the case in all models investigated. 

We obtain the physical parameters that best describe the SED of \PKS using the minimisation routines of \jetset which utilises {\scshape{Minuit}}. The relevant quantities include the doppler factor of the emitting region, $\delta$, magnetic field strength, $B^{\prime}$, size and location of the emitting region with respect to the base of the jet $r_{b}$ and $r_{\rm diss}$ respectively, which we detail in the following section. We assume throughout that $\delta = \Gamma$, with $\Gamma$ the Lorentz factor of the motion in the jet of \PKS. 
Using the values of these quantities determined by the leptonic fit of the SED,
we calculate the corresponding neutrino emission of \PKS under additional assumptions about the proton content of the jet, in a second step, using a semi-analytical approach, as described in~\cite{Oikonomou:2019djc}. 

Briefly, we assume that the relativistic protons in the jet are accelerated to a power law with the same spectral
index as the spectral index of the electron population determined by the minimisation
performed with \jetset in the jet of \PKS. Protons are accelerated up to a maximum energy determined by assuming that the acceleration timescale $t'_{\rm acc}$, is similar to the proton-energy-loss timescale in the acceleration zone, $t'_{\rm cool}$, given by,

\begin{equation} 
t_{\rm acc}^{'-1} > t_{\rm cool}^{'-1} \equiv t_{\rm cross}^{'-1} + t_{p,
  {\rm syn}}^{'-1} + t_{p \gamma}^{'-1},
\end{equation}
where the synchrotron cooling time for protons with energy $\varepsilon^{\prime}_p$ in a magnetic field with strength $B$
is given by,
\be
t'_{p,{\rm syn}} = 6 \pi m_p^4c^3/(m_e^2 \sigma_T {B^{\prime}}2
\varepsilon^{\prime}_p),\ee
where $m_p$ and $m_e$ are the proton and electron mass respectively, $c$ the speed of light, and $\sigma_T$ the Thomson cross section. The blob crossing time, $t'_{\rm cross} = r_{b}/c$, is used to approximate the adiabatic energy loss rate. 

\noindent Here, 
\be
t'_{\rm acc} = \varepsilon'_p/(\eta ceB'),
\label{eq:tacc}
\ee
is the proton acceleration timescale. The
acceleration efficiency is parametrised by $\eta \leq 1$. We consider several  values for $\eta$ in the range $\eta = 10^{-3} - 1$. The latter corresponds to the fastest  possible acceleration, and can be achieved in the Bohm limit, in diffusive shock acceleration. The lower values of $\eta$ are more conservative and the lower range considered here is inferred for blazars in certain studies~\cite{Inoue:1996vv,Dermer:2014vaa,Inoue:2016fwn}. 
\noindent The timescale for photomeson interactions is estimated as,
\begin{equation} 
t_{p\gamma}^{'-1} = \frac{c}{2 \gamma^{\prime 2}_p}
\int^{\infty}_{\bar{\varepsilon}_{\rm th}} {\rm d}
\bar{\varepsilon'_{\gamma}} \sigma_{p \gamma} \kappa_{p \gamma}
\bar{\varepsilon'_{\gamma}}
\int^{\infty}_{\bar{\varepsilon'_{\gamma}}/(2\gamma^{\prime}_p)} {\rm d}
\varepsilon'_{t} \varepsilon^{'-2}_{t} 
  n'_{\gamma},
\label{eq:pgammaRate}
\end{equation}
\noindent where, $\bar{\varepsilon}_{\rm th} \sim 145$~MeV is the threshold
photon energy for photopion production in the proton rest frame, and
$\sigma_{p \gamma}$ and $\kappa_{p \gamma}$ are the cross section and
inelasticity of photomeson interactions, respectively. We use the parametrisations of~\cite{Murase:2005hy} for $\sigma_{p \gamma}$ and
$\kappa_{p \gamma}$. The quantity
$n'_{\gamma}$ is the target photon density of photons with energy $\vareps^{\prime}_t$ differential in
energy and $\gamma^{\prime}_p$ is the Lorentz factor of the proton. 
The fraction of energy converted to pions is given by,
$f_{p \gamma} \equiv t'_{\rm cool} / t'_{\rm p\gamma}$.

The all-flavour neutrino luminosity per logarithmic energy produced by \PKS in the conditions outlined above, is given by
\be
\vareps_{\nu}L_{\vareps_{\nu}} \approx \frac{3}{8} f_{p\gamma}(\vareps_{p}) \vareps_{p}L_{\vareps_{p}},
\label{eq:nuLuminosity}
\ee
where the proton luminosity per logarithmic energy $\vareps_{p}L_{\vareps_{p}}$ is a free parameter constrained only loosely by the SED of \PKS, and the neutrinos emerge at energy $\vareps_{\nu} \sim 0.05 \vareps_p$. 

 Using the procedure detailed in Appendix~\ref{sec:cascade} we then calculate the additional radiation produced due to the interactions of pions and protons inside the source. Accounting for any additional interactions of this cascade emission before escaping the source, we determine, in a third step, the maximum proton luminosity compatible with a particular SED. In other words, we make sure that the combined radiation produced by the accelerated electrons and protons in the emission region does not overshoot the observed SED of \PKS. In practice, we define the bolometric proton luminosity $L_p$ as a multiple of the bolometric photon luminosity $L_{\gamma}$, with $L_p = \xi_{\rm cr} L_{\gamma}$, where the multiplicative factor $\xi_{\rm cr}$ is typically referred to as the baryon loading factor. 

To find the maximum proton luminosity compatible with the SED for a particular combination of source parameters, we start with the assumption of a low proton luminosity and increase it until the \chisq of the leptonic-only fit starts to increase, due to the contribution of cascade radiation. Here and throughout by \chisq, we will denote the \chisq of the leptonic-only fit unless otherwise stated. The goodness of fit for a particular proton luminosity is quantified by the $\chi^2_{\rm total}$ of the combined (hadronic and leptonic) fit, 
$\chi^2_{\rm total} = \chi^2 + \delta \chi^2.$
We define the maximum proton luminosity, and thus the maximum baryon loading, $\xi_{\rm cr,max}$ consistent with a particular SED as the value which corresponds to $\delta\chi^2 = 4$, in other words, a $2\sigma$ increase. 

\subsection{Synchrotron-self-Compton model} 
\label{subsec:ssc_modelling} 
We first studied whether the quiet SED of \PKS can be well described within the synchrotron-self-Compton (SSC) model. The observed spectral shape requires that the electron spectrum steepens with increasing energy. We assumed that the underlying emitting electron population has a power-law spectrum, where the density of electrons with Lorentz factor $\gamma^{\prime}_e$ is given by,  
\be
N^{\prime}(\gamma^{\prime}_e) = N_{0} \Theta(\gamma^{\prime}_e - \gamma^{\prime}_{e,\rm min}) \gamma^{s_e} \mathrm{e}^{-\gamma^{\prime}_e/\gamma^{\prime}_{e,\rm cut}},
\ee

\noindent where $N_{0}$ is the normalisation of the electron number density, $\gamma^{\prime}_e$, $\gamma^{\prime}_{e,\rm min}$, and $\gamma^{\prime}_{e,\rm cut}$, are the electron Lorentz factor, the minimum assumed $\gamma^{\prime}_e$, and the value of $\gamma^{\prime}_e$ beyond which the density of the electrons drops exponentially, accordingly, and $s_e < 0$ the power-law index of the electron spectrum. We performed a six-parameter fit using {\scshape{Minuit}}.
We scanned the parameter space in terms of the Doppler factor, $\delta$ and the magnetic field strength, $B^{\prime}$, while allowing {\scshape{Minuit}} to optimise the remaining parameters, namely the radius of the emitting zone, $r_{b}$, $s_e$, $\gamma^{\prime}_{\rm cut}$, and $N_0$. For the former we scanned in the range $\delta = [5, 50]$ and for the latter, in the range $\log(B^{\prime}/\rm G) = [-3, -1.3]$. For the scan we moved along the grid for which we considered 18 bins in $\log B^{\prime}$ and $46$ bins in $\delta$. 
\begin{table}[h]
\begin{center} 
\caption{Parameter values assumed fixed in the SSC model investigated for the long-term SED of \PKS.
\label{tab:param_ssc}}
\begin{tabular}{lll}
Parameter & Symbol/Units & Value \\
\hline
Minimum electron Lorentz factor & $\gamma^\prime_{e, \min}$ & 10 \\
Minimum proton Lorentz factor & $\gamma^\prime_{p, \min}$ & 10 \\
Proton spectral index & $s_p$ & $s_{e}$ \\
Jet dissipation radius & $r_{\rm diss}$ (cm) & $\delta r_b$ \\
\end{tabular}
\end{center} 
\end{table}

For the remaining leptonic parameters, the range was left free. We fixed $\gamma^{\prime}_{e,\rm min} = 10$. We did not attempt to fit the Planck data, since we know from the analysis of~\cite{Karamanavis_b} that the radio and \gRay data are not co-spatially produced. 

Once the leptonic fits are obtained, we introduce an additional proton component. The proton spectrum is assumed to be a power-law with minimum Lorenz factor $\gamma^{\prime}_{p, \rm min} = 10$, spectral index $s_p = s_e$, i.e. fixed from the results of the leptonic fit, and $\gamma^{\prime}_{p,\rm max}$ determined from setting $t^{\prime}_{\rm acc} = t^{\prime}_{\rm cross}$, where for the former we use eq.~(\ref{eq:tacc}) and a range of values for $\eta$ which parametrised the acceleration efficiency in the source. The fixed parameters of the SSC models studied are summarised in table~\ref{tab:param_ssc}. 

\subsection{External Compton on Dust Torus}
\label{subsec:DT_modelling}

We next investigated whether the SED of \PKS can be well described as originating from a single emission zone inside the DT (but beyond the BLR). We assumed that the DT intercepts $50\%$ of the accretion disk radiation, and re-emits it in the infrared. In this model, the high-energy (X-ray and \gRay) emission is expected to be predominantly external-Compton (EC) emission from the interactions of electrons with the photons of the torus. Hence, we refer to this model as EC-DT. 

\begin{table}
\begin{center}
\caption{Parameter values assumed fixed in the EC-DT model investigated for the long-term SED of \PKS.
\label{tab:param}}
\begin{tabular}{lll}
Parameter & Symbol/Units & Value \\
\hline
Minimum electron Lorentz factor & $\gamma^\prime_{e, \min}$ & 1 \\
Minimum proton Lorentz factor & $\gamma^\prime_{p, \min}$ & 10 \\
Proton spectral index & $s_p$ & $s_{e,1}$ \\
Jet dissipation radius & $r_{\rm diss}$ (cm) & $\delta r_b$ \\
Accretion disk temperature & $T_{d}$ (K) & $10^4$ \\
Dust torus temperature & $T_{\rm DT}$ (K)  & 370 \\
Dust torus radius & $r_{\rm DT}$ (cm) & $8\times 10^{18}$ \\
Accretion disk luminosity & $L_{d}$ (erg s$^{-1}$) & $1 \times 10^{46}$ \\
Fraction of $L_{d}$ intercepted by DT & $f_{\rm DT}$ & 0.5 \\
\end{tabular}
\end{center} 
\end{table}

The spectral shape of the DT was approximated by a black-body spectrum that has a rest-frame peak temperature of  $T = 370$~K following~\cite{Ghisellini_2009}. Interestingly, the WISE data exhibit a kink at this frequency, but the DT emission cannot be responsible for explaining the kink, even under the extreme assumption that it intercepts and re-radiates the majority of the disk emission.

 We fixed the peak temperature of the accretion disk emission to $T_{\rm AD}$ = 10000~K and the spectral shape was approximated by a black-body spectrum. Since in this scenario the accretion disk is behind the emitting region it does not play an important role in our results.

The spectrum of electrons was approximated by a broken power law, with indices $s_{e,1}$ and $s_{e,2}$ below and beyond the break energy $\gamma^{\prime}_{e, \rm break} m_e c^2$. We performed a seven-parameter ($r_b$, $B$, $\delta$, $s_{e,1}$, $s_{e,2}$, $\gamma^{\prime}_{e, \rm break}$, $\gamma^{\prime}_{e, \rm max}$) scan and present the results in the $\log B^{\prime}-\delta$ plane in figure~\ref{fig:DT_scan}. The location of the emitting region was assumed to be $r_{\rm diss} = \delta r_b$. 
Additional parameters relating to the spectrum, and size of the dust torus were kept fixed and are reported in table~\ref{tab:param}. A coarse initial scan in $\log B^{\prime}$ led to successful fits in the range $\log(B^{\prime}/\rm G) = [- 0.5, - 0.4]$. In what follows we focus on this range
in which we performed a fine scan (40 bins in $\log B^{\prime}$). For the Doppler factor, we scanned in the range $\delta = 10 - 50$ with a total of 41 bins. 
\
The remaining leptonic parameters were left free, except for $\gamma^{\prime}_{e,\rm min}$ for which we tested two values, namely, $\gamma^{\prime}_{e,\rm min} = 1, 100$. As in the SSC case, we did not include the Planck data in the fit. 

For the proton spectrum we followed the same procedure as in the SSC scenario assuming that the spectral index follows that of the electron spectrum below the break, i.e. $s_{e,1}$. We tested several values of the parameter $\eta$.

\begin{figure}
\begin{center} 
\includegraphics[width = 0.65  \textwidth, clip, rviewport=0.05 0.075 0.95 0.95]{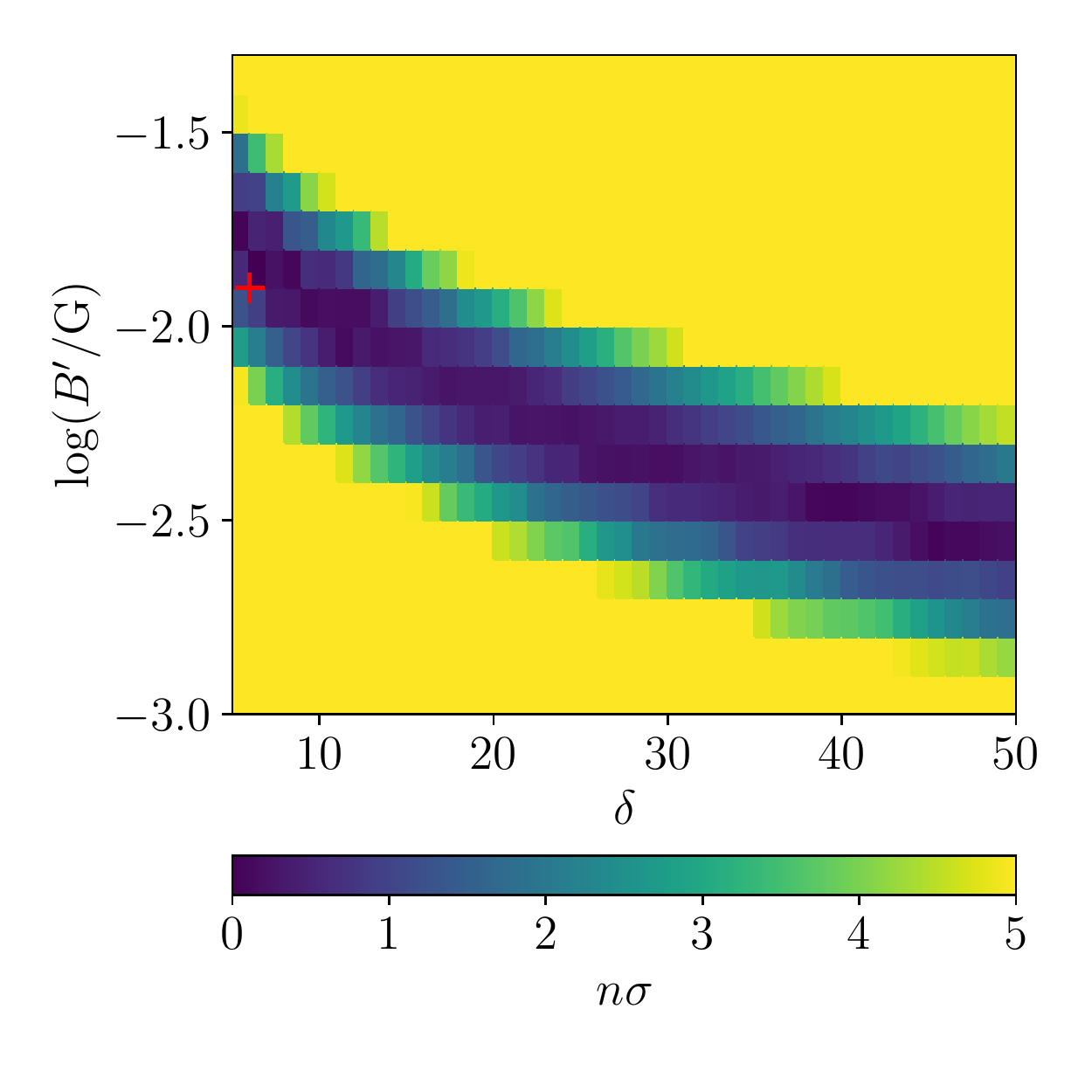}
\caption{Systematic uncertainty on the determination of $\delta$ and $\log B^{\prime}$ ($B^{\prime}$ in Gauss) of the quiet SED of \PKS considering only synchrotron and SSC emission. The colormap gives the departure from the best-fit parameters in units of n$\sigma$ (see main text for details). The red cross denotes the best-fit parameters.}
\label{fig:SSC_nSigma} 
\end{center} 
\end{figure}

\section{Results}
\label{sec:results}
\subsection{Synchrotron-self-Compton model}
\label{subsec:ssc} 

We present the results of the six parameter scan in the $\log B^{\prime}$ vs $\delta$ plane. Figure~\ref{fig:SSC_nSigma} gives the deviation of the fitted SED for each combination of $\log B^{\prime}$ and $\delta$ with respect to the best-fit SED in units of, $n \sigma = S \sqrt{\chi^2 - \chi^2_\text{min}},$ with $\chi_{\rm min}^2$ the $\chi^2$ of the best-fit realisation (see also~\cite{Oikonomou:2019djc}). The scale factor $S~=~1/\sqrt{\chi^2_\text{min}/\text{ndf}}$, 
where ndf the number of degrees of freedom, is an approximate correction~\cite{Rosenfeld:1975fy} employed to enlarge the uncertainty due to having a poor minimum $\chi^2$. This may either signify a wrong model or simplified model assumptions, or underestimated uncertainties in the data. In our case, the non fully simultaneous nature of the data in the different wavelength bands, plays a role in leading to high $\chi_{\rm min}^2$ values.

We find a long and shallow minimum in the $\log B^{\prime}$ vs. $\delta$ plane. Since presumably the true minimum \chisq lies outside the scan range, we present the significance scan using the $\chi^2_\text{min}$ shown with a red cross (i.e. the minimum within the fitting range considered). The long shallow trend is consistent with expectations from the SSC model, as the ratio of the peak frequencies translates to a constant value of $B^{\prime} \delta$, whereas the ratio of the peak luminosities translates to an upper limit on the product of $B\delta^3$ (see e.g. Eqs. 4 and 11 of~\cite{Tavecchio:1998xw} and discussion therein), and additional observational constraints are needed to identify unique values. The corresponding range of fitted blob sizes inside the $1\sigma$ contour is $r_b \sim 3\times10^{17} - 8\times10^{18}$~cm and the electron spectal index is $s_e = 1.6 - 1.7$. 

\begin{figure*}  
\includegraphics[width = \linewidth]{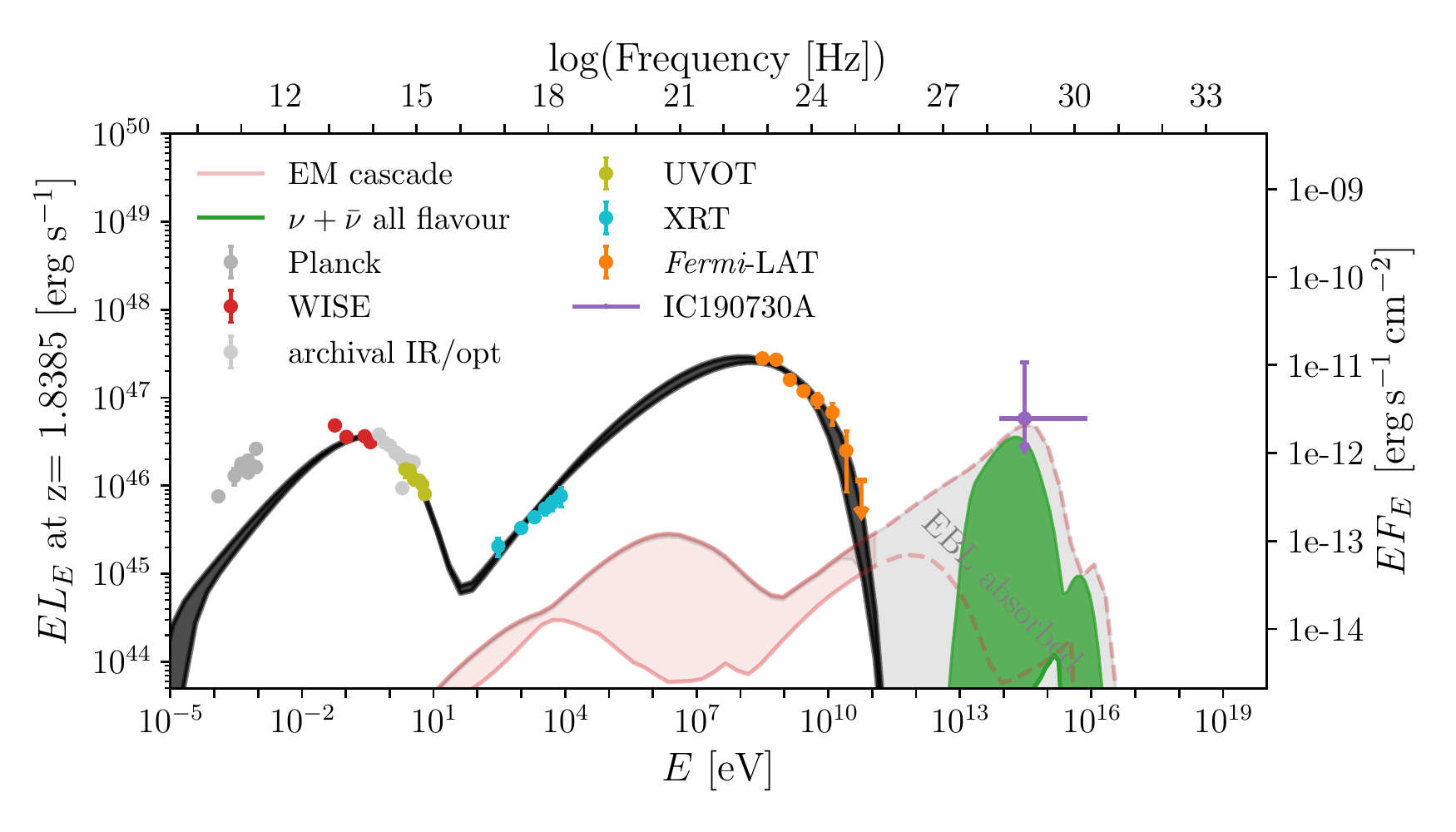}
\caption{The spectral energy distribution of \PKS and SSC fit to the SED. Grey datapoints are {\it Planck} and archival IR/optical data not included in the fit. WISE, archival IR/opt and {\it Planck} datapoints were obtained from the SSDC online database. UVOT and XRT datapoints are from this analysis, see section~\ref{subsec:data} for details. The \Fermi-LAT datapoints are from the analysis of~\cite{franckowiak2020patterns} (quiet epoch). The mean all-flavour neutrino flux corresponding to the observation of 1 muon neutrino for an assumed live time of ten years is shown in purple. The 90\% CL upper limit is also shown (assuming $N_{\rm background} \approx 0$). The fits shown correspond to those that are within $1\sigma$ from the best-fit parameters as shown in Figure~\ref{fig:SSC_nSigma}. The black shaded region shows the total expected photon SED. The neutrino flux expected from interactions of protons with the internal radiation field are shown in green. The pink shaded region shows the accompanying electromagnetic cascade emission.}
\label{fig:SSC_SED_low_eta}
\end{figure*}

\begin{figure}  
\includegraphics[width = \linewidth]{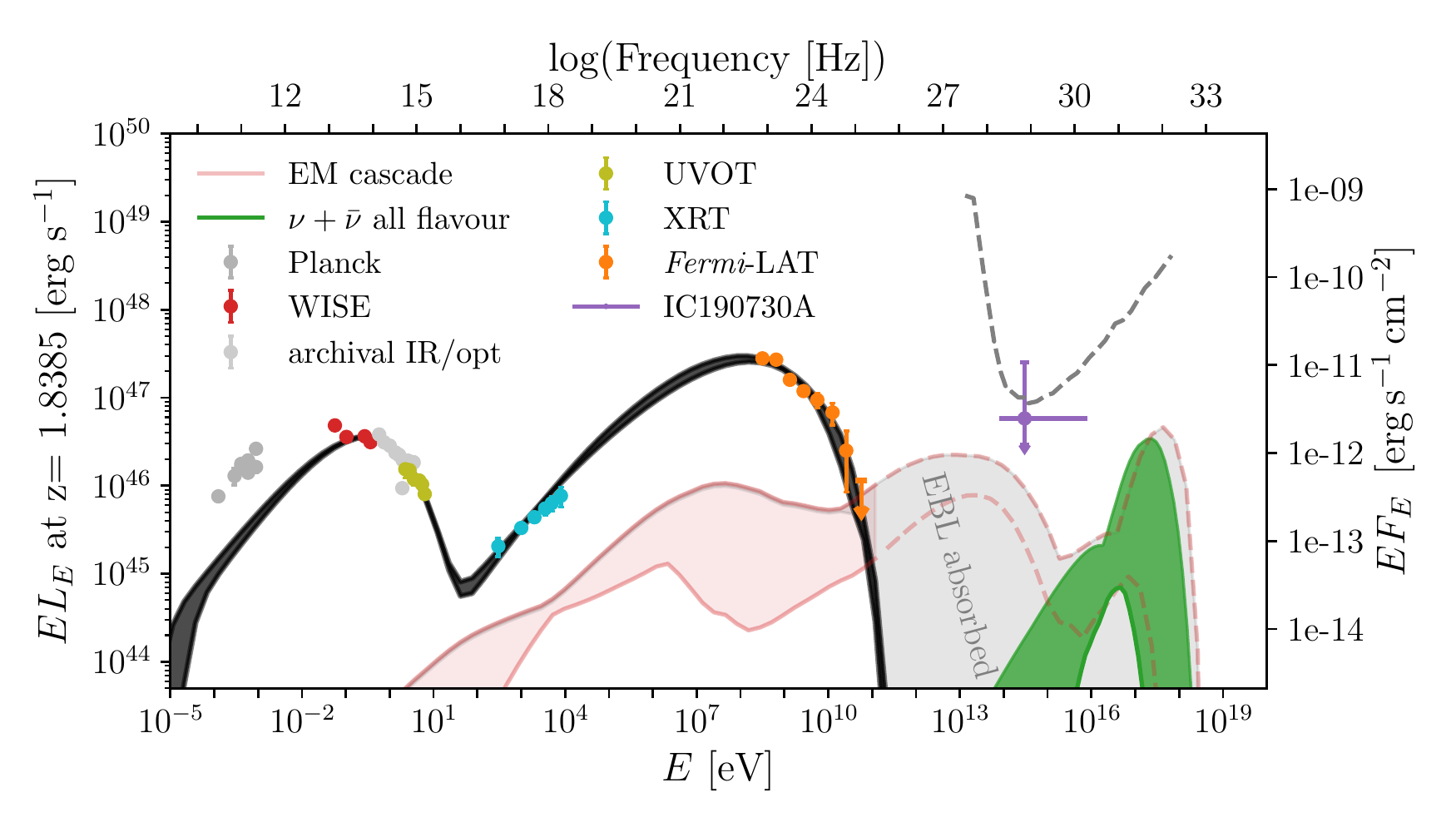}
\caption{Same as Figure~\ref{fig:SSC_SED_low_eta}, but with a higher value of maximum proton energy, parametrised by acceleration efficiency (see eq.~\ref{eq:tacc}) $\eta = 0.1$. The differential sensitivity of the GFU analysis at the declination of \PKS is also shown here as $3\vareps_{\nu}/[A_{\rm eff}(\vareps_{\nu},\delta) \ln 10]$ by the dashed grey curve, such that following the curve for one order of magnitude in energy yields one neutrino event for a $\mathrm{d}N/\mathrm{d} \vareps_{\nu} \sim \vareps_{\nu}^{-2}$ power-law neutrino spectrum. For details about the observational data see the caption of Figure~\ref{fig:SSC_SED_low_eta}.}
\label{fig:SSC_SED_high_eta}
\end{figure}

The SEDs that fall within $1\sigma$ from the minimum $\chi^2$ are shown in Fig~\ref{fig:SSC_SED_low_eta}. The black band shows the expected combined leptonic and hadronic energy flux after accounting for expected attenuation by the EBL. The expected neutrino energy flux associated with this model is shown in green, under the assumption that protons are accelerated in the jet with the maximum luminosity allowed by the SED. The associated cascade emission is shown in pink. The interactions of photons and electrons of hadronic origin inside the source environment have been taken into account as detailed in Appendix~\ref{sec:cascade}. The electromagnetic cascade emission at energies beyond $\sim 0.1$~TeV is absorbed by interactions with the intervening EBL and shaded in grey. 

The results of figure~\ref{fig:SSC_SED_low_eta} have been obtained under the assumption of a low acceleration
efficiency, where $\eta = 10^{-3}$. Due to this choice, the maximum proton energy is of order 
$\varepsilon_{p,max} \sim 3\times 10^{16}~\mathrm{eV} (\eta/10^{-3}) (r_b/10^{18}) (B/0.01~\mathrm{G}) (\delta/10) \approx 10^{16}-10^{17}$~eV. 

Multiplying the expected neutrino flux differential in energy with the average GFU effective area in the declination range $\delta = [-5^{\circ}, 30^{\circ}]$ we find that the most probable neutrino energy in this model is $\sim 200-300$~TeV, comparable to the most probable energy quoted by IceCube for IC-190730A assuming an $\vareps_{\nu}^{-2.19}$ neutrino spectrum. The second (smaller) peak in the $1\sigma$ band of neutrino spectra (at higher energy) corresponds to a subset of models with larger doppler factors. 

The same results but assuming $\eta = 0.1$ are shown in 
figure~\ref{fig:SSC_SED_high_eta}. A larger value of $\eta$ shifts the peak of the neutrino spectrum to higher energies, and away from the energy range where IceCube expects the majority of neutrinos at the declination 
of \PKS in the GFU analysis.  

\begin{figure*}
\includegraphics[width = 0.32 \linewidth, clip,rviewport=0.05 0 0.95 1]{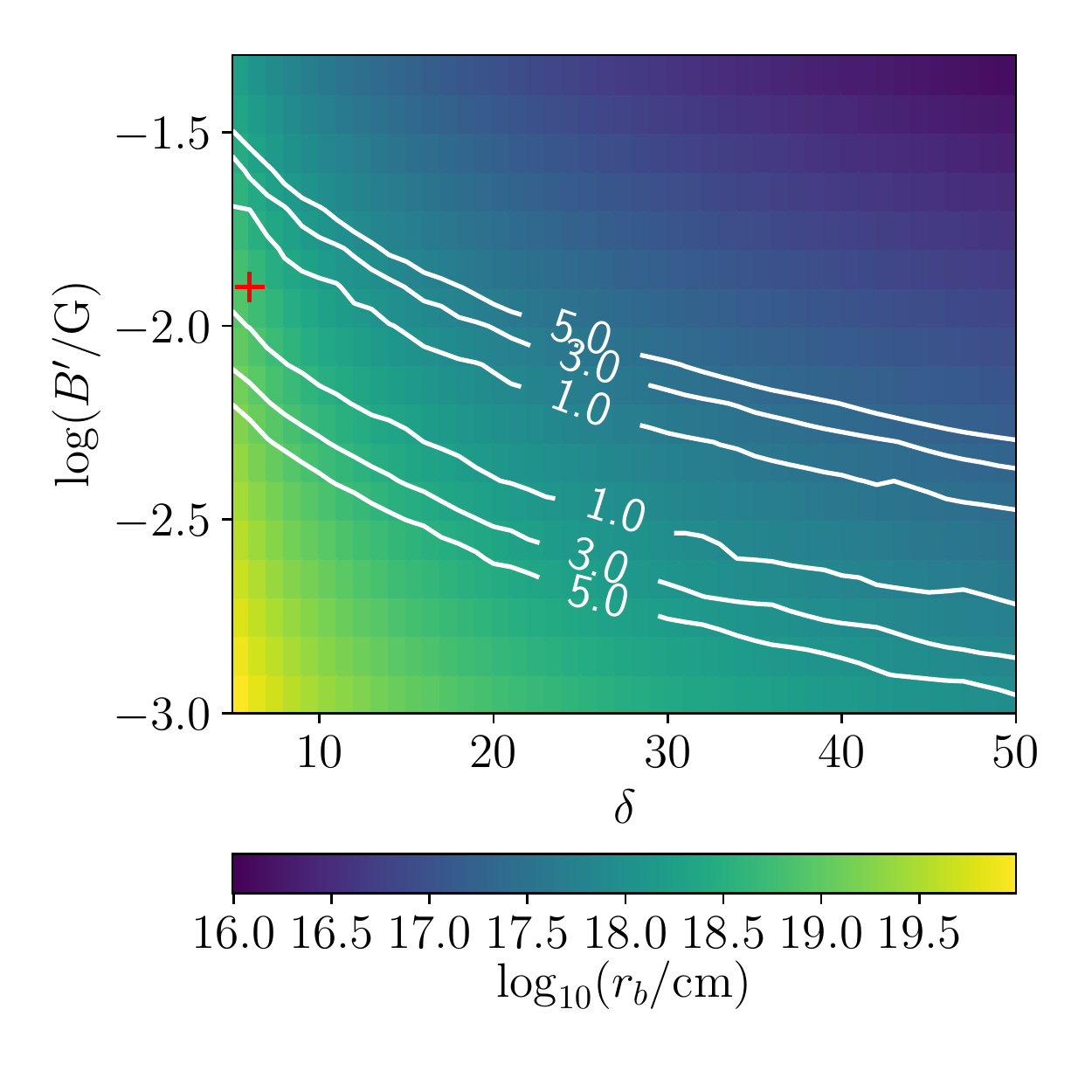}
\includegraphics[width = 0.32 \linewidth, clip,rviewport=0.05 0 0.95 1]{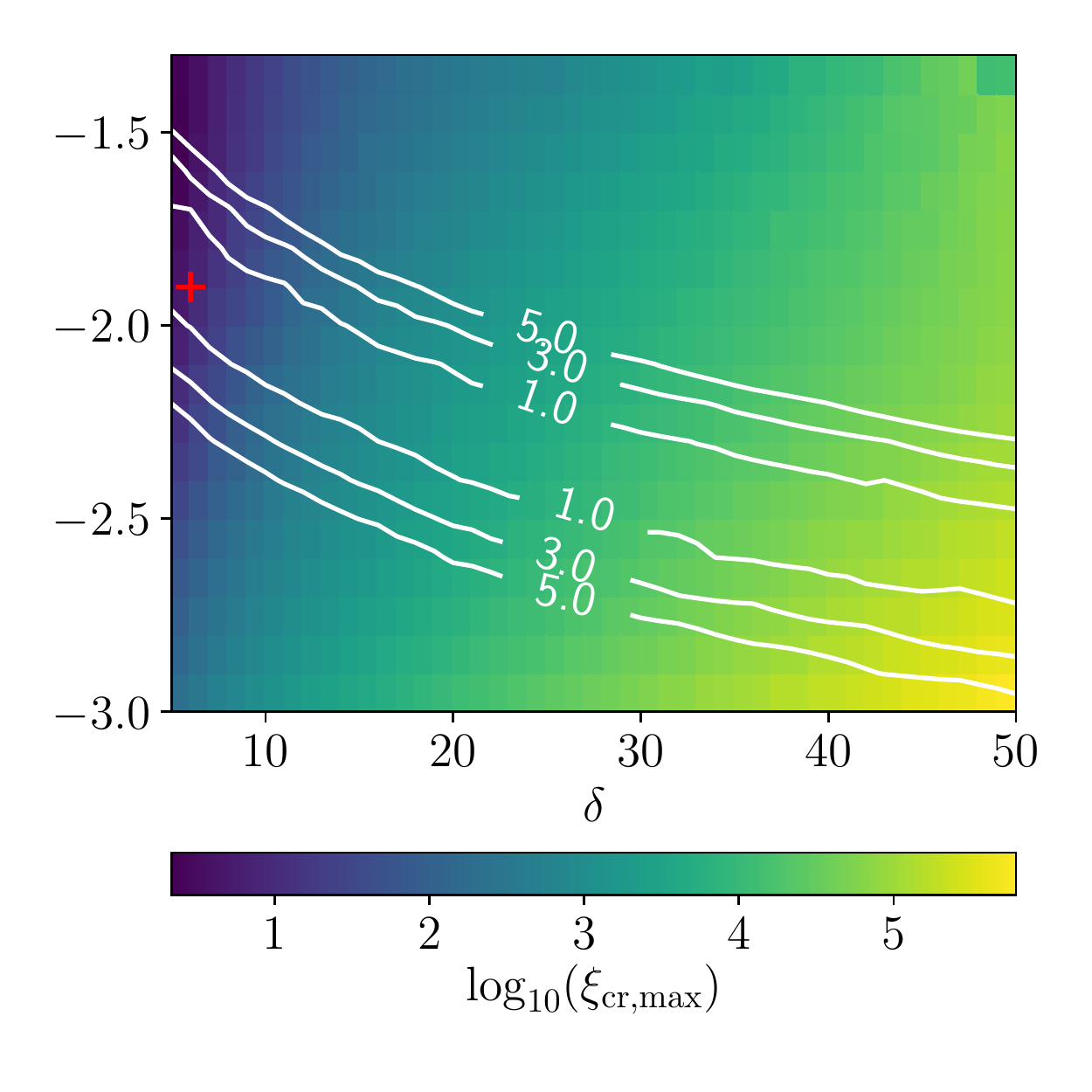}
\includegraphics[width = 0.32 \linewidth, clip,rviewport=0.05 0 0.955 1]{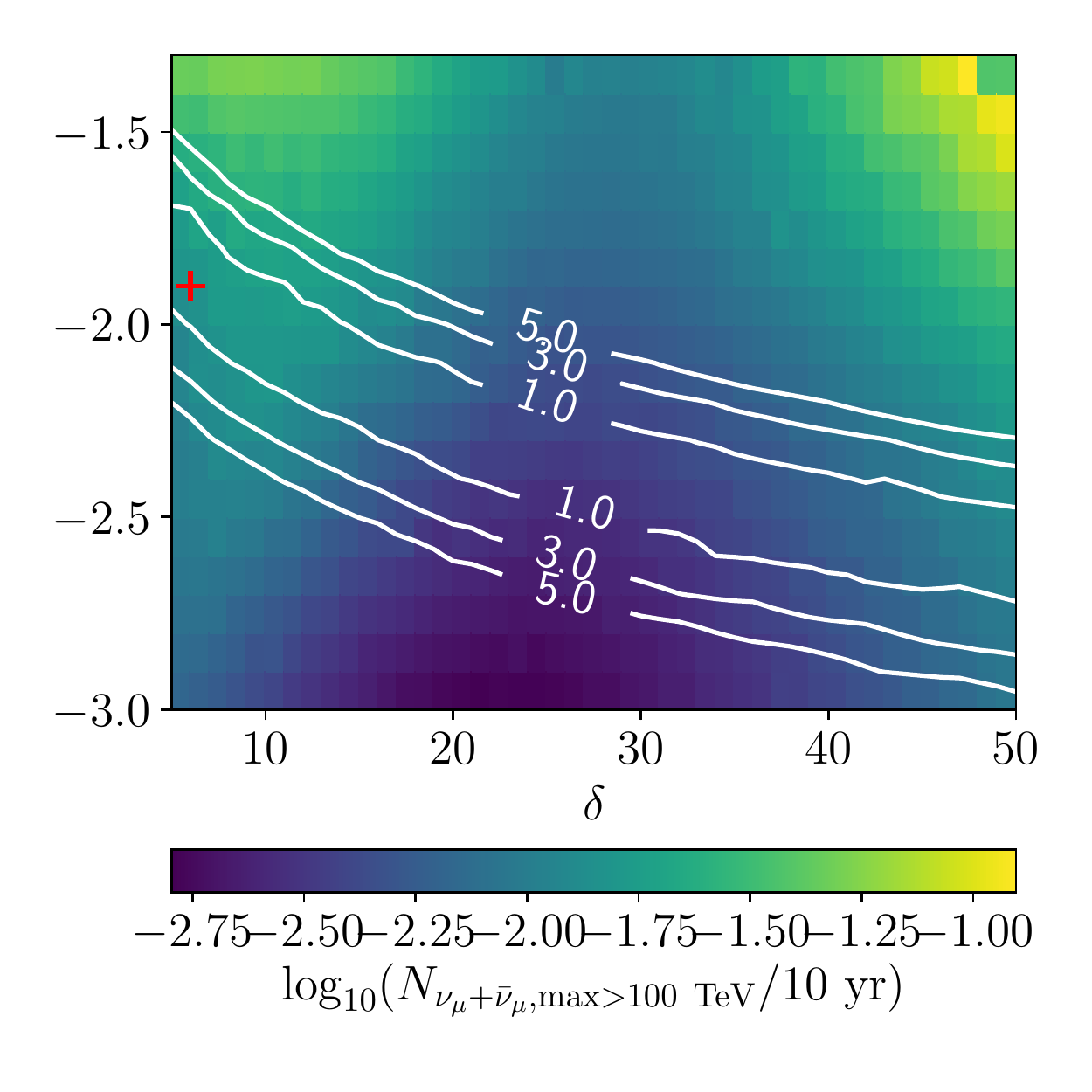}
\caption{The fitted radius of the emitting region (left), maximum baryon loading factor (middle), and resulting expected number of muon neutrinos (right) in the scanning range considered in this work in the SSC scenario. The number of muon neutrinos expected has been calculated under the assumption of constant neutrino luminosity during the ten years covered. The value of the acceleration efficiency parameter is assumed to be $\eta = 0.1$ here. The contours indicate the one, three and five sigma regions from the minimum $\chi^2$ of the leptonic fit which is indicated with a red cross.}
\label{fig:SSC_heatmaps}
\end{figure*}

Since the observed luminosity is fixed by the observations and the luminosity of the synchrotron and self-Compton humps is approximately $L_{{\rm syn}(C)} \sim 4\pi r_b^2 c \delta^4 U_{B\rm(rad)}$, with $U_B$ and $U_{\rm rad}$ the energy density of the magnetic field and of the synchrotron radiation respectively, a larger value of $\delta$ results in lower values of $r_b$. This can be seen on the left panel of figure~\ref{fig:SSC_heatmaps}. 

The right panel of figure~\ref{fig:SSC_heatmaps} shows the expected number of muon neutrinos in the entire scan range when $\eta = 0.1$.\footnote{Qualitatively the results are similar for $\eta = 10^{-3}$, but we consider the model with $\eta = 0.1$ more realistic as it corresponds to more modest proton luminosity as we discuss next.} The corresponding maximum baryon loading factor allowed by the fit to the SED as a result of the electromagnetic cascade emission, $\xi_{\rm cr,max}$, is shown in the middle panel. A much higher value of $\xi_{\rm cr, max}$ is observed with increasing $\delta$. This is because the total allowed cascade luminosity is fixed by the data, and so is the total observed luminosity $L_{\rm Tot, obs} = L_{p, \rm casc} + L_{\gamma}$, where $L_{\gamma}$ is the total observed luminosity of the leptonic SED. The proton-induced cascade luminosity is roughly proportional to $L_{p, \rm casc} \sim L_{\gamma} \xi_{\rm cr, max} n^{\prime}_{\gamma} r_b \sigma_{\rm eff}$, where, for simplicity, we don't specify which process and denote the cross section with $\sigma_{\rm eff}$ which could apply to photomeson or photopair processes. The comoving photon number density,  $n_{\gamma}^{\prime} \sim L_{\gamma}/(\delta^4 4 \pi r_b^2 c \vareps^{\prime}_{\gamma})$, strongly decreases with increasing $\delta$ (despite the fact that the decreasing $r_b$ slightly counteracts the effect). As a result, a much larger value of $\xi_{\rm cr, max}$ is allowed by the fit with increasing $\delta$. 
In general, for the values of $\eta$ we investigated, the total neutrino luminosity follows the same increasing trend with $\delta$ as $\xi_{\rm cr, max}$. The combination of smaller $\vareps_{p,\rm max}$ due to a smaller blob radius, and larger proton luminosity result in a larger neutrino luminosity. Additionally, for the particular value of $\eta$ shown here as seen on the right panel of figure~\ref{fig:SSC_heatmaps} an enhanced neutrino flux can be seen for low values of $\delta$. This is because at low values of $\delta$ the cascade emission (higher $\vareps_{p,\rm max}$) shifts to higher energies, alleviating the constraint posed by the XRT data to some extent and allowing for additional cascade emission. An increase of about four orders of magnitude of the observed proton luminosity is accompanied by about two orders of magnitude variation in the expected number of neutrinos. The reason why the expected number of neutrinos does not grow as fast as $\xi_{\rm cr, max}$ with increasing $\delta$ is that the neutrino production efficiency is larger for smaller values of $\delta$, owing to the higher density of target photons ($n^{\prime}_{\gamma} \propto \delta^{-4} r_b^{-2}$). 

The number of muon and antimuon neutrinos with energy above 100 TeV expected with this model in the GFU channel with ten years of exposure is given in table~\ref{tab:nu_rate_ssc}, under the assumption that the emitted neutrino flux of \PKS was constant during this entire period. The highest number of neutrinos is expected for $\eta \sim 10^{-3}$, with $\mathcal{N}_{\nu_{\mu}+\bar{\nu}_{\mu}}(> 100 ~\rm TeV) = 0.1\pm 0.1$. 
For the higher values of $\eta$ investigated the number of neutrino counts expected is almost an order of magnitude lower, because neutrino production shifts to higher energy, away from the peak sensitivity of IceCube (see for example the grey dashed line in figure~\ref{fig:SSC_SED_high_eta} which gives the differential sensitivity at declination $\sim 10^{\circ}$). 
 table~\ref{tab:nu_rate_ssc} also shows the neutrino flux expected during the time spanned by the \Fermi-LAT light curve of figure~\ref{fig:lightcurve}, under the assumption that the neutrino luminosity of \PKS at time $t$, $L_{\nu}(t)$, is given by, 
\be 
L_{\nu}(t) = \left(\frac{\phi_{\Fermi}(t)}{\phi_{\Fermi}(t_0)}\right)^{2} L_{\nu}(t_0)
\ee 
where $\phi_{\Fermi}(t)$ is the \Fermi-LAT flux of \PKS in the 100 MeV-800 GeV energy range at time $t$, and $\phi_{\Fermi}(t_0)$ the flux of \PKS at the time of arrival of \icnu. Here, $L_{\nu}(t_0)$ is the neutrino luminosity expected in the SSC model at the time of the quiet SED of \PKS which we use throughout this work as representative of the SED at the time of arrival of \icnu. The quadratic dependence of $L_{\nu}$ on $\phi_{\Fermi}$ can be understood as follows. 
$L_{\nu} \propto L_p n^{\prime}_{\gamma} \propto L_p L_{\gamma} \propto L_{\gamma} L_{\gamma} \propto \phi_{\Fermi}^2$, where $n^{\prime}_{\gamma} \propto L_{\gamma}/\delta^4 r_b^2 \vareps_{\gamma}^{\prime} $ is the number density of target photons available for photomeson interactions which is directly proportional to $L_{\gamma}$ and we assume as previously that the proton luminosity is proportional to the photon luminosity of the source. Similar to $L_{\nu}$ the luminosity of secondary leptons from photomeson interactions also scales as $\phi_{\Fermi}^2$. We also show the fraction of the flux expected during the quiet epoch (MJD 55266-57022) in table~\ref{tab:nu_rate_ssc}. The table shows that if the neutrino luminosity of \PKS has a quadratic dependence on the \Fermi-LAT flux, the expected neutrino signal is strongly enhanced in the SSC model, but the majority of neutrinos are expected from times when the source was at a high \gRay state, and not at quiet times such as the time of arrival of \icnu.

The maximum proton luminosity allowed by the fit which was used to obtain the expected number of neutrinos above, $L_{p, \rm max} = \xi_{cr, \rm max} L_{\gamma}$, can be compared to the Eddington luminosity of the super-massive black hole of \PKS. 
The black hole mass of \PKS is $M_{\rm BH}\sim 10^9 M_{\odot}$, which corresponds to Eddington luminosity $L_{\rm Edd} = 1.26\times10^{47}$\ergs~\cite{10.1046/j.1365-8711.2003.06255.x,2006ApJ...637..669L}. 
For a comparison of these two quantities we can convert the maximum proton luminosity in the observer frame, $L_{p, \rm max}$ to the absolute, beaming corrected proton luminosity, $\mathcal{L}_{p, \rm max} = L_{p,\rm max} / (2\Gamma^2)$. The range spanned by the scenarios described in this section is given in table~\ref{tab:nu_rate_ssc}. 

For the value of $\eta$ which gives the maximum expected neutrino events, we find that the ratio $\mathcal{L}_p/L_{\rm Edd}$ is in the range $10 - 10^5$. Though jet emission with super-Eddington power is not alarming, especially for short periods, see e.g. ~\cite{10.1093/mnras/stv1802}, models with a large ratio $\mathcal{L}_p/L_{\rm Edd}$ present an energetic challenge and are thus disfavoured. Lower ratios of $\mathcal{L}_p/L_{\rm Edd}$ are obtained for larger values of $\eta$, but at the cost of reduced expected neutrino flux. 

\begin{table}
\begin{center} 
\caption{Muon and antimuon neutrinos expected to be detected by IceCube (GFU analysis) for parameter sets within $1\sigma$ from the best-fit scenario in the SSC model for different values of the acceleration efficiency parameter, $\eta$. We show a scenario in which the neutrino flux is constant for the entire ten years of IceCube live time ($L_{\nu} \propto L_{\gamma}^0$), and a scenario in which the neutrino flux scales with the \gRay luminosity of the source ($L_{\nu} \propto L_{\gamma}^{2}$) during the time spanned by the \Fermi-LAT light curve (MJD 54684-58695), $T_{\Fermi~\rm  lc}$. In the latter case we also show the contribution of the quiet period (MJD 55266-57022) only, $T_{\rm quiet}$. The range of values of the proton luminosity in terms of the Eddington luminosity of \PKS spanned by 68\% of scenarios closest to the $\chi^2_{\rm min}$, $\log_{10}(\mathcal{L}_p/L_{\rm Edd})$, is also given. 
\label{tab:nu_rate_ssc}}
\begin{tabular}{ccccc}
$\eta$ & \multicolumn{3}{c}{$\mathcal{N}_{\nu_{\mu}+\bar{\nu}_{\mu}}(> 100 ~\rm TeV)$} & $\log_{10}(\mathcal{L}_p/L_{\rm Edd})$\\
\hline
& $L_{\nu} \propto L_{\gamma}^0$ & \multicolumn{2}{c}{$L_{\nu} \propto L_{\gamma}^{2}$} & \\
& 10yr & $T_{\Fermi \rm ~lc}$ & $T_{\rm quiet}$ & \\
\hline
0.001 & $0.1\pm0.1$ & $3\pm3$ & $0.2^{+0.4}_{-0.2}$ & [1, 5] \\
0.01 & $0.03^{+0.1}_{-0.01}$ & $0.9^{+3}_{-0.8}$ & $0.02^{+0.03}_{-0.01}$ & [1, 4]\\
0.1  & $0.01^{+0.01}_{-0.005}$ & $0.3^{+0.3}_{-0.2}$ & $0.05^{+0.2}_{-0.05}$ & [0, 3] \\
\end{tabular}
\end{center} 
\end{table}

Scenarios within the $1\sigma$ contour in this model, additionally suffer from too large derived blob size. For the SSC emitting region not to be larger than the opening angle of a conical jet, it should be located at $r_{\rm diss} \gtrsim \delta r_b = 4 - 13$~pc. The results of the radio observations of \PKS place the 86~GHz core at $4.1\pm0.4$~pc from the jet base~\cite{Karamanavis_b}, but the \gRay emitting region at $\sim 2$~pc. 
As a result, the range of radii predicted in this model would be expected to result in variability on timescales of order $t_{\rm var, obs} \sim r_b / [\delta c (1+z)]$ be 6 - 10 days for $\delta \geq 40$ and larger for smaller values of $\delta$, which is much larger than inferred from the \Fermi-LAT analysis of e.g.~\cite{2021MNRAS.503.3145B}. 

 The simultaneous fit to the XRT and \Fermi-LAT data is poor as can be seen in Figs.~\ref{fig:SSC_SED_low_eta}-\ref{fig:SSC_SED_high_eta}, as a single slope cannot perfectly connect the two datasets. This observation is true even prior to the addition of a proton (and thus cascade) component. This makes the SSC model difficult to reconcile with the SED of \PKS and suggests the existence of an additional, external-Compton, emission component. It should be kept in mind that the XRT data do not span the entire period covered by the \Fermi-LAT data. But unless the XRT average is harder than what was captured by the six observation epochs, a better agreement is not expected. Our results are consistent with the results of~\cite{Ciprini:2009kr}, who reached similar conclusions as to the applicability of the SSC model for the post-2008-flare SED of \PKS.  

If we exclude the WISE data from the leptonic fit, we find solutions with a smaller blob radius $r_b = 10^{16} - 10^{17}$~cm, implying a distance $r_{\rm diss} \geq \delta r_b = 0.1-1$~pc from the base of the jet. This circumvents the problem that the SSC fit suggests a  since the smaller emitting region can in principle be well upstream of the derived jet distance of the radio data, and allows for emitting regions small enough to be consistent with variability timescales of $\sim 1$~day. However, the poor simultaneous fit to the XRT-\Fermi-LAT data remains.  

\subsection{External Compton on Dust Torus}
\label{subsec:DT_result}

We next present the results of the leptonic scan under the assumption that the emitting region of \PKS is located beyond the BLR but within the dust torus. Here, we have assumed that $\gamma^{\prime}_{e,\rm min} = 1$. 

\begin{figure}
\begin{center} 
\includegraphics[width = 0.65 \linewidth, clip,rviewport=0.05 0.075 0.95 0.95]{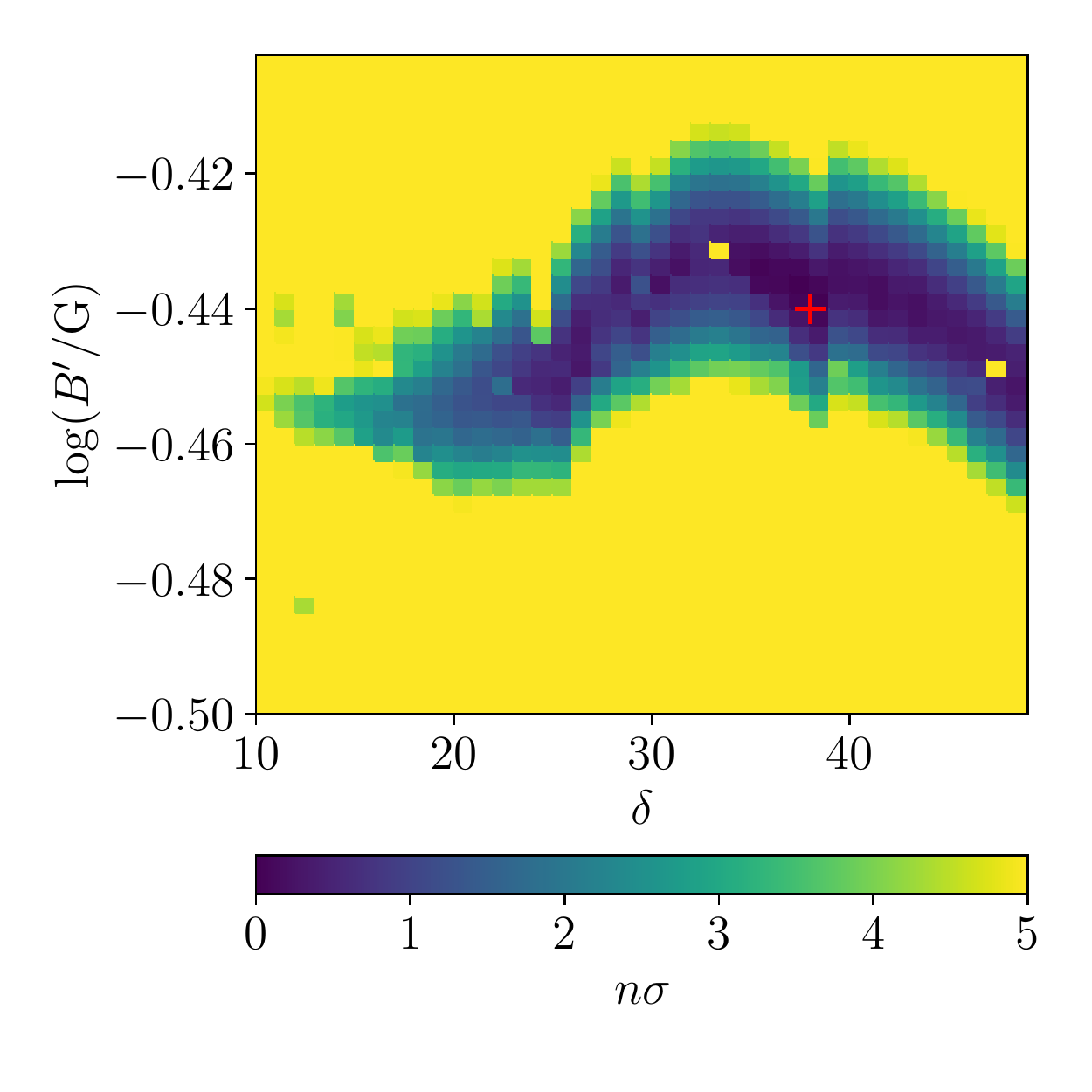}
\caption{Systematic uncertainty on the determination of $\delta$ and $\log B^{\prime}$ ($B^{\prime}$ in Gauss) of the quiet SED of \PKS. Same as Figure~\ref{fig:SSC_nSigma}, but considering additionally external Compton emission on photons from the dust torus. The colormap gives the departure from the best-fit parameters in units of n$\sigma$. The red cross denotes the best-fit parameters. 
\label{fig:DT_scan}}
\end{center} 
\end{figure}

The goodness of fit of the studied scenarios within the scanning range is shown in figure~\ref{fig:DT_scan}. We find a minimum that is deep in $B^{\prime}$ but long and shallow in $\delta$. All solutions within 1$\sigma$ from the best-fit are in the range $B^{\prime} = 0.34-0.38$~G. Interestingly, this is consistent with the magnetic field value derived from long-term radio and optical observations in a completely independent manner by~\cite{Shao:2020tth}. 

On the other hand, $\delta$ can take values between 20-50 without a significant change of the $\chi^2$ of the fit. The best-fit $\chi^2$ is significantly better than the minimum found with the SSC scenario. 

\begin{figure}
\begin{center}
\includegraphics[width = 0.65 \linewidth]{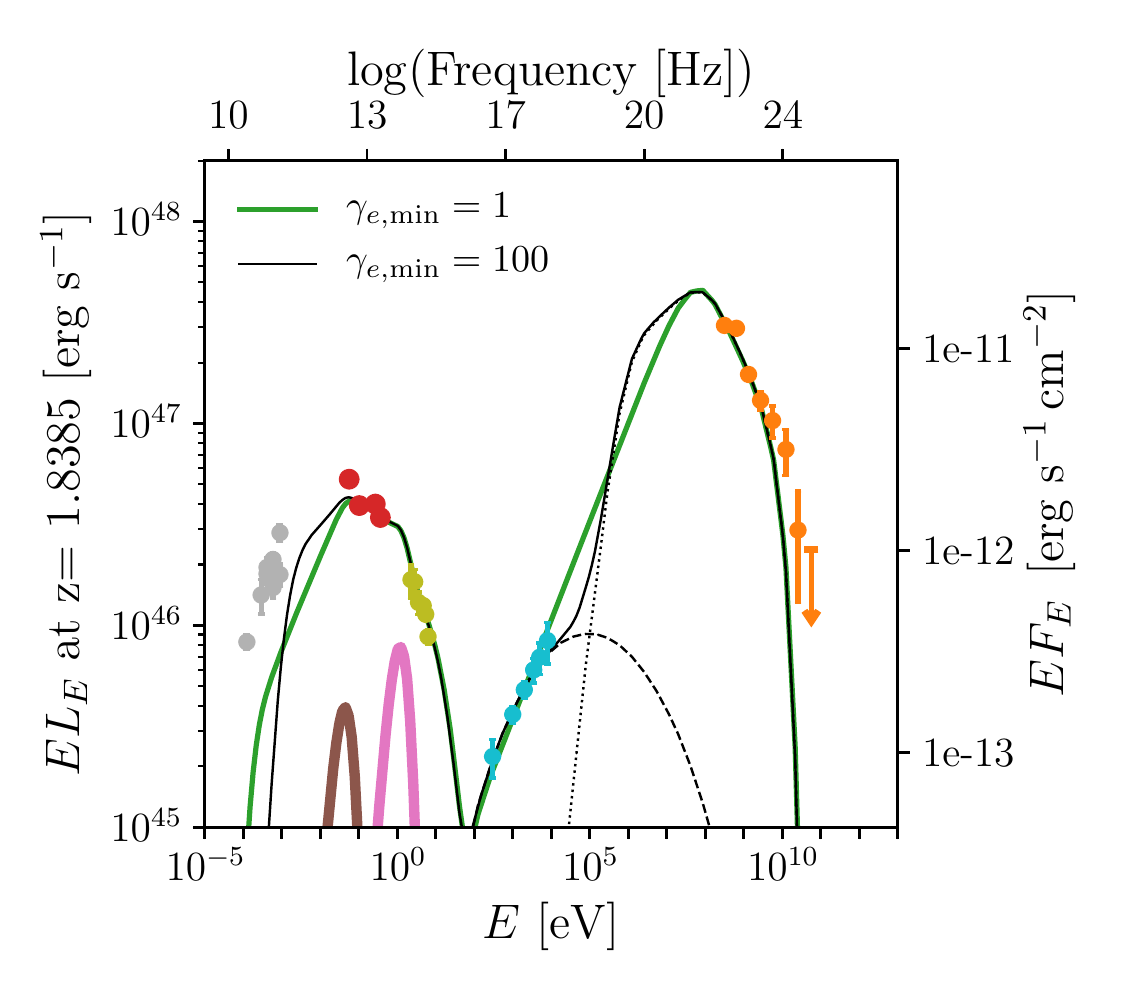}
\caption{Example fits to the long-term SED of \PKS close to the minimum $\chi^2$ with the EC-DT model, for two different values of $\gamma^{\prime}_{e,\rm min}$. The assumed disk emission is shown in pink, and the assumed DT emission is shown in brown. The case with assumed $\gamma^{\prime}_{e,\rm min} = 100$ is shown in black. The contribution of the SSC component is shown by the dashed line, the dotted line shows the EC-DT emission only, and the total emission is shown by the solid line. The case with $\gamma^{\prime}_{e,\rm min} = 1$ is shown in green. Here the contribution from SSC is below the plotting range. Details about the observational datapoints are given in the caption of figure~\ref{fig:SSC_SED_low_eta}.}
\label{fig:DT_SEDs}
\end{center} 
\end{figure}

We find an additional family of solutions with larger $\gamma^{\prime}_{e, \rm min} = 100$, which result in a very similar $\chi^2$. An example close to the minimum for $\gamma^{\prime}_{e, \rm min} = 1, 100$ is shown in figure~\ref{fig:DT_SEDs}. The main difference between the two models is that in the case of larger $\gamma^{\prime}_{e,\rm min}$ the XRT data are produced by SSC emission, and the EC emission starts to dominate at higher energy. However, from the point of view of parameters relevant for neutrino production in our formalism, the two scenarios yield almost identical results as they point to an emitting region of the same size, and the $\log B^{\prime}$ and $\delta$ constraints we obtain by scanning are very similar. Therefore we only show the expected neutrino spectra for the case with $\gamma^{\prime}_{e, \rm min} = 1$ in what follows. 

\begin{figure*}
\includegraphics[width = \linewidth]{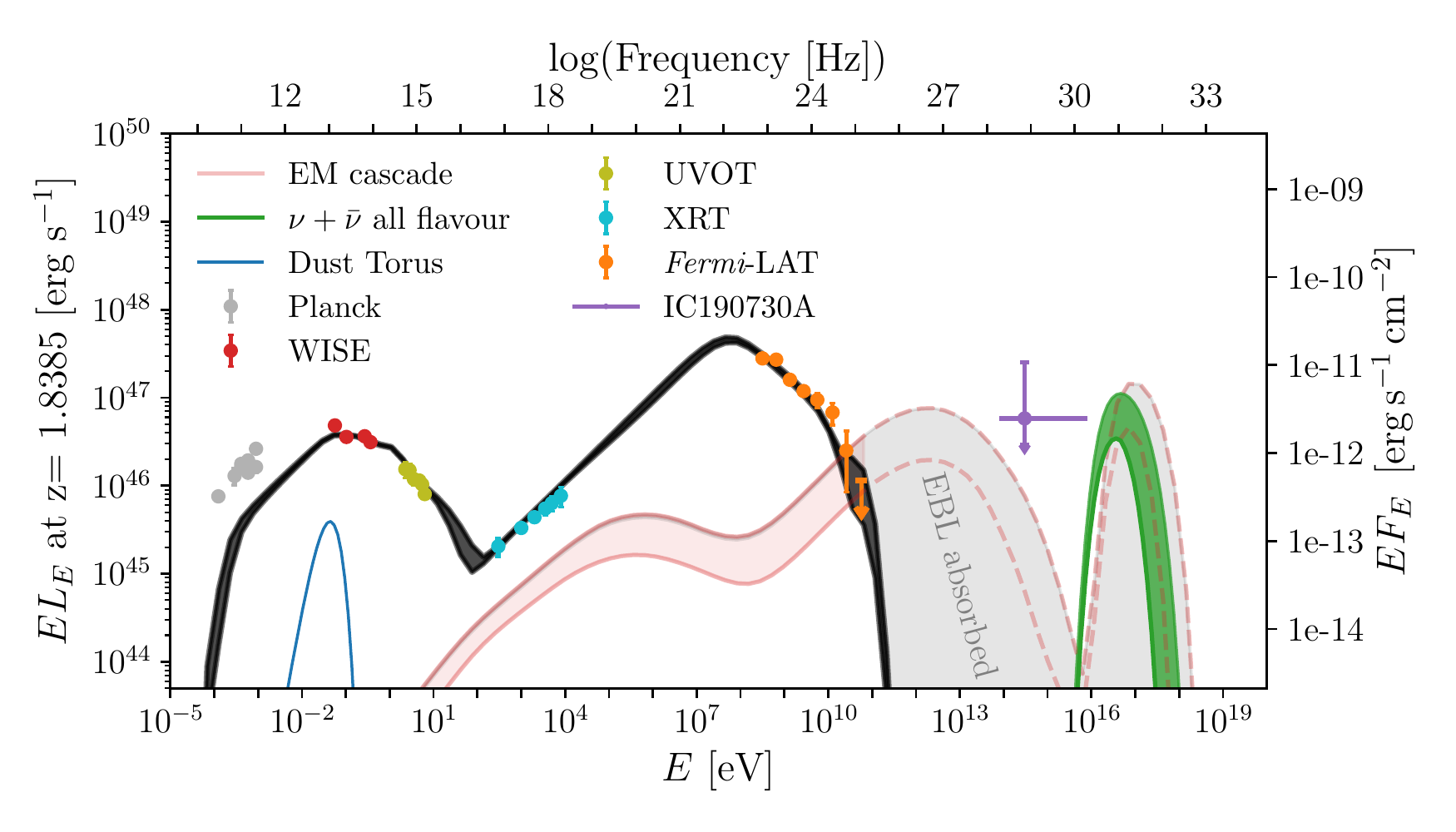}
\caption{The model SEDs and associated neutrino energy spectra for parameter combinations within $1\sigma$ from the best-fit scenario indicated with a red cross in Figure~\ref{fig:DT_scan}. The black shaded region shows the total expected photon SED, while the blue solid line shows the assumed spectrum of the dust torus. The sum of neutrinos expected from interactions of protons with the internal/blob and DT radiation field are shown in green. The pink shaded region shows the accompanying electromagnetic cascade emission. The mean all-flavour neutrino flux, corresponding to the observation of 1 muon neutrino for an assumed live time of ten years is shown in purple. The 90\% CL upper limit is also shown (assuming $N_{\rm background} \approx 0$). Details about the observational datapoints are given in Figure~\ref{fig:SSC_SED_low_eta}.}
\label{fig:DT_sed_and_neutrinos}
\end{figure*}

The neutrino spectra shown in figure~\ref{fig:DT_sed_and_neutrinos} correspond to parameter combinations within 1$\sigma$ from the best-fit in the numerical scan, assuming $\eta = 10^{-2}$. We have not included the contribution of neutrinos from interactions with deboosted BLR/disk photons here, because we found it to be negligible. In the models shown in figure~\ref{fig:DT_sed_and_neutrinos} the maximum neutrino luminosity is limited by the electromagnetic cascade emission whose maximum we have obtained in the same way as in the SSC scenario, as described in section~\ref{sec:model}.

We considered several values of the parameter $\eta$ but we show the case of $\eta = 0.01$ in the plots, which yielded the maximum expected neutrino counts, which we report in table~\ref{tab:dt_nu_rate} (obtained using eq.~~\ref{eq:Nnu}). We found that for larger values of $\eta$ the cascade luminosity and neutrino luminosity increase, but without any gain in terms of the expected number of neutrinos as the peak of the neutrino flux shifts to higher energies, which are not detectable with the IceCube through-going muon analysis, while increased cascade luminosity limits $\xi_{\rm cr, max}$ to lower values. For smaller values of $\eta \lesssim 0.01$, the photopion production efficiency does not reach its maximum on the infrared photons of the dust torus. The above comment reflects the general trend. On the other hand, within the $1\sigma$ region the three values of $\eta$ we report on in table~\ref{tab:dt_nu_rate} are compatible within the corresponding error bars. For the most favourable value of $\eta = 0.01$, the neutrino flux can be such as to expect at most, $\mathcal{N}_{\nu_{\mu}+\bar{\nu}_{\mu},\mathrm{\rm max
}}^{> 100 ~\rm TeV} \approx 0.1\pm 0.1$
muon and antimuon neutrinos with energy exceeding 100 TeV to have been detected by the IceCube GFU analysis during the 10~year operation of IceCube, in the 1$\sigma$ range of models considered assuming that the neutrino luminosity of \PKS is assumed constant with time. The probability to observe one neutrino at any time during the IceCube 10-year live time is thus $\sim 10\%$.

\begin{table}
\begin{center}
\caption{Muon and antimuon neutrinos expected to be detected by IceCube with ten years of exposure, with the IceCube GFU effective area for parameter sets within $1\sigma$ from the best-fit scenario in the EC-DT models studied.  We show a scenario in which the neutrino flux is constant for the entire ten years of IceCube livetime ($L_{\nu} \propto L_{\gamma}^0$), and a scenario in which the neutrino flux scales with the \gRay luminosity of the source ($L_{\nu} \propto L_{\gamma}^{1.5}$) during the time spanned by the \Fermi-LAT lightcurve (MJD 54684-58695), $T_{\Fermi~\rm  lc}$. In the latter case we also show the contribution of the quiet period (MJD 55266-57022) only, $T_{\rm quiet}$. The range of values of the proton luminosity in terms of the Eddington luminosity of \PKS spanned by 68\% of scenarios closest to the $\chi^2_{\rm min}$, $\log_{10}(\mathcal{L}_p/L_{\rm Edd})$ is also given. \label{tab:dt_nu_rate}}
\begin{tabular}{ccccc}
$\eta$ & \multicolumn{3}{c}{$\mathcal{N}_{\nu_{\mu}+\bar{\nu}_{\mu}}(> 100 ~\rm TeV)$} & $\log_{10}(\mathcal{L}_p/L_{\rm Edd})$ \\
\hline
& $L_{\nu} \propto L_{\gamma}^0$ & \multicolumn{2}{c}{$L_{\nu} \propto L_{\gamma}^{1.5}$} & \\
& 10yr & $T_{\Fermi \rm ~lc}$ & $T_{\rm quiet}$ & \\
\hline
0.001 & $0.03^{+0.04}_{-0.03}$ & $0.3^{+0.4}_{-0.3}$ & $0.03^{+0.04}_{-0.03}$ & [-2, -1]]\\
0.01 & $0.1^{+0.05}_{-0.06}$ & $0.9\pm0.5$ & $0.1^{+0.06}_{-0.05}$ & [-2, -1]\\
0.1 & $0.06^{+0.04}_{-0.03}$ & $0.7^{+0.3}_{-0.4}$ & $0.07\pm 0.03$ & [-2, -1] \\ 
\end{tabular}
\end{center} 
\end{table}

Table~\ref{tab:dt_nu_rate} also shows the neutrino flux expected during the time spanned by the \Fermi-LAT light curve of figure~\ref{fig:lightcurve}, under the assumption that the neutrino luminosity of \PKS at time $t$, $L_{\nu}(t)$, is given by, 
\be 
L_{\nu}(t) = \left(\frac{\phi_{\Fermi}(t)}{\phi_{\Fermi}(t_0)}\right)^{1.5} L_{\nu}(t_0)
\ee 
where $\phi_{\Fermi}(t)$ is the \Fermi-LAT flux of \PKS in the 100 MeV-800 GeV energy range at time $t$, and $\phi_{\Fermi}(t_0)$ the flux of \PKS at the time of arrival of \icnu. The dependence of $L_{\nu}$ on $\phi_{\Fermi}$ can be understood as follows. 
$L_{\nu} \propto L_p f_{p\gamma} \propto L_p n^{\prime}_{\rm DT} r_{\rm DT} \propto L_{\gamma} L_{\gamma}^{1/2} \propto \phi_{\Fermi}^{1.5}$, where $n^{\prime}_{\rm DT} \propto \Gamma^2 L_{\rm DT}/r_{\rm DT}^2$ is the number density of target photons from the DT available for photomeson interactions which is approximately constant, and $r_{\rm DT}\propto L_{d}^{1/2} \propto L_{\gamma}^{1/2}$ where in the last step we made the reasonable assumption that the disk luminosity is proportional to the \gRay luminosity of the source and we further assumed, as throughout, that the proton luminosity is proportional to the photon luminosity of the source. Similar to $L_{\nu}$ the luminosity of secondary leptons from photomeson interactions also scales as $\phi_{\Fermi}^{1.5}$. We also show the fraction of the neutrino counts expected during the quiet epoch MJD 55266-57022, $T_{\rm quiet}$ in table~\ref{tab:dt_nu_rate}. Similar to what was observed in table~\ref{tab:nu_rate_ssc}, if the neutrino luminosity of \PKS follows $L_{\nu}\propto L_{\gamma}^{1.5}$, the expected neutrino signal is strongly enhanced, but the majority (90\%) of neutrinos are expected from times when the source was at a high state, and only 10\% of neutrinos are expected to arrive during the quiet \gRay epoch (which spans, according to our definition, approximately $50\%$ of the last ten years).

\begin{figure*}
\includegraphics[width = 0.32 \linewidth, clip,rviewport=0.05 0 0.95 1] {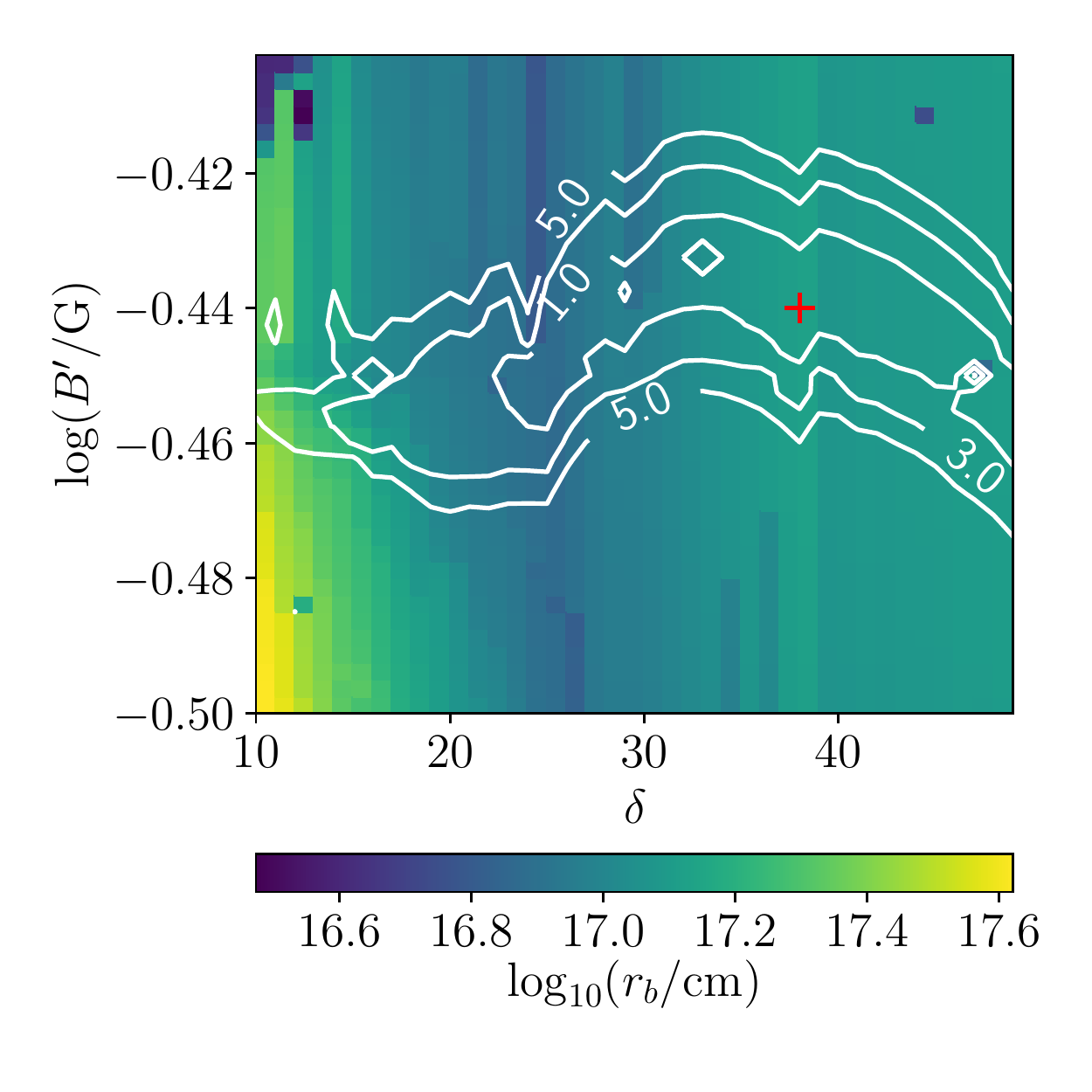}
\includegraphics[width = 0.32 \linewidth,  clip,rviewport=0.05 0 0.95 1]{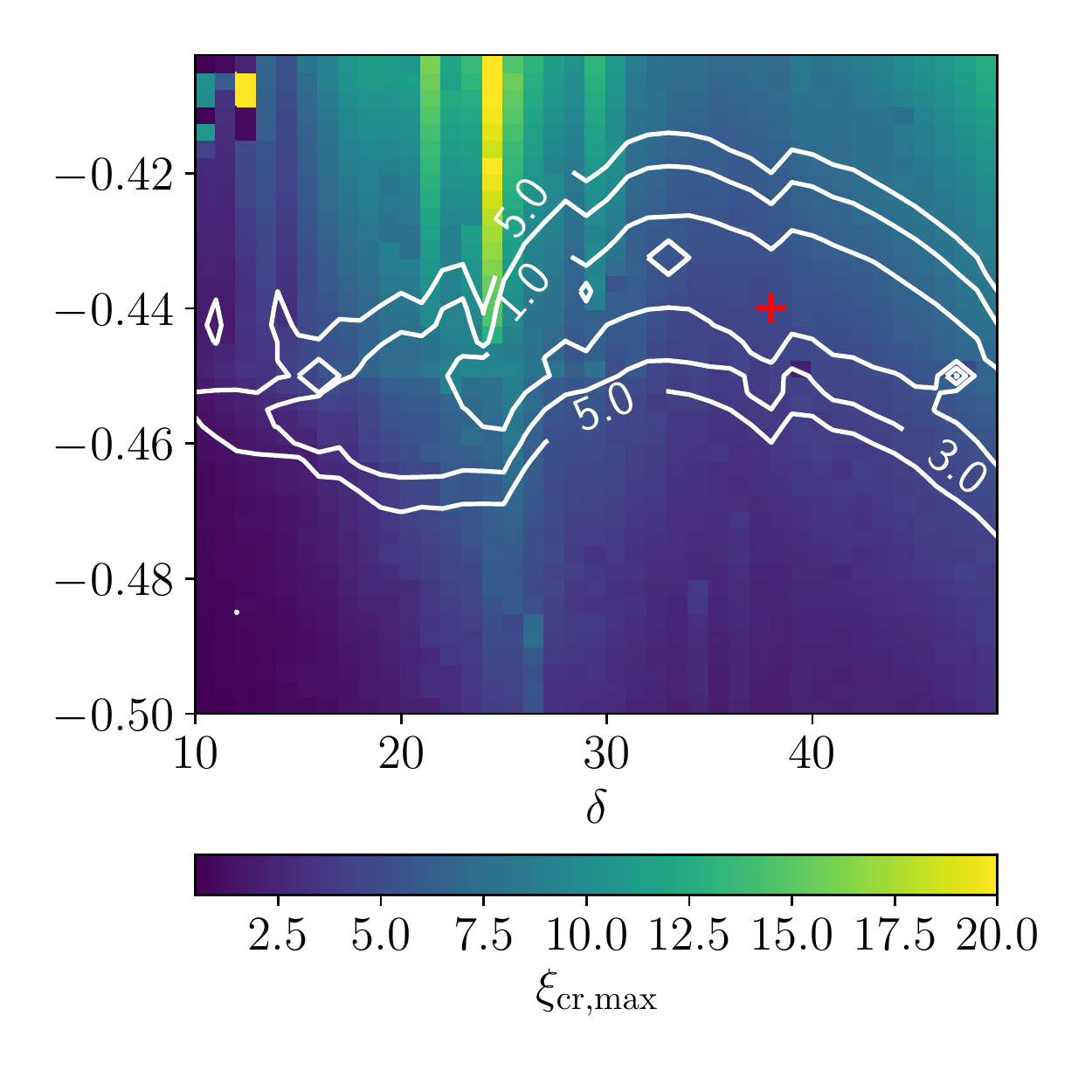}
\includegraphics[width = 0.32 \linewidth,  clip,rviewport=0.05 0 0.95 1]{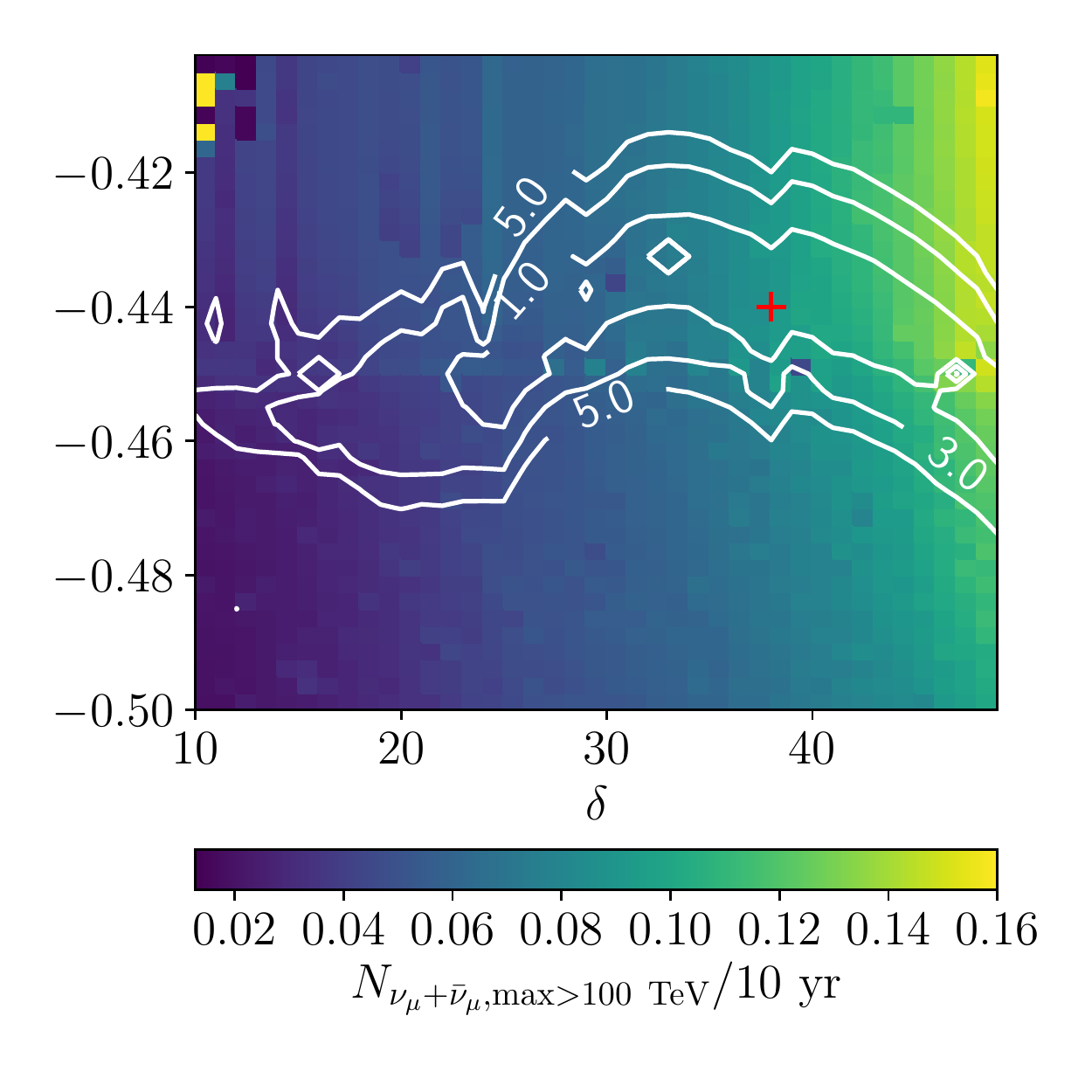}
\caption{The fitted radius of the emitting region (left), maximum baryon loading factor (middle), and resulting expected number of muon neutrinos (right) in the scanning range considered in this work in the EC-DT scenario. The number of muon neutrinos expected has been calculated under the assumption of constant neutrino luminosity during the ten years covered. The value of the acceleration efficiency parameter is assumed to be $\eta = 0.01$ here. The contours indicate the one, three and five sigma regions from the minimum $\chi^2$ of the leptonic fit which is indicated with a red cross.}
\label{fig:DT_heatmaps}
\end{figure*}

The neutrino counts expected assuming constant neutrino luminosity are shown for the entire scanning range on the right-hand panel of figure~\ref{fig:DT_heatmaps}. The maximum baryon loading factor is shown in the middle panel. The baryon loading factor is almost constant in the entire scanning range whereas the expected neutrino luminosity and expected neutrino counts increase by a factor of $\sim 10$ with increasing $\delta$. 

The total observed luminosity $L_{\rm Tot, obs}$ is equal to the leptonic only observed luminosity $L_{\gamma}$ and the total cascade luminosity $L_{p, \rm casc,abs}$, where the subscript {\it abs} has been added to denote the effect of EBL absorption which is taken into account here. In practice we have, $L_{\rm Tot, obs} = L_{\gamma} + L_{p, \rm casc,abs} \approx L_{\gamma}$ since the cascade luminosity is much smaller than $L_{\gamma}.$ The observed luminosity $L_{\rm Tot, obs}$ is fixed by the observational data. The cascade luminosity depends on $\xi_{\rm cr, max}$ as, $L_{p, \rm casc}\sim \xi_{\rm cr, max} L_{\gamma} n^{\prime}_{\rm DT} r_b \sigma_{\rm eff}$, where $n^{\prime}_{\rm DT}$ is the number density of DT photons in the comoving frame, and $\sigma_{\rm eff}$ is the cross section of the process which may be photomeson or photopair interactions.  

The rest-frame luminosity of the torus, $L_{\rm DT}$, its characteristic (rest-frame) energy,
$\vareps_{\rm DT}$, and radius, $r_{\rm DT}$ are fixed. The comoving number density of DT photons is thus given by, $n^{\prime}_{\rm DT} \sim L_{\rm DT} 
\delta^2 / (4\pi r_{\rm DT}^2 \vareps_{\rm DT}^{\prime}) \sim L_{\rm DT} \delta^2 / (4\pi r_{\rm DT}^2 \vareps_{\rm DT} \delta)$, meaning that $n^{\prime}_{\rm DT} \propto \mathrm{constants} \cdot \delta$. Thus, we can write the cascade luminosity as, $L_{\rm casc} \sim \mathrm{constants} \cdot \xi_{\rm cr} L_{\gamma} \delta r_b \sigma_{\rm eff} $. The cascade luminosity increases with $\delta$. This can also be seen in figure~\ref{fig:DT_sed_and_neutrinos}, where the lower part of the confidence band corresponds to low values of $\delta$ and larger values of $\delta$ correspond to moving upwards (higher $L_{\rm casc}$) on the band. On the one hand, fits with larger $\delta$ result in a synchrotron cascade spectrum which rises with a harder slope in the XRT energy range, thus effectively the fit can allow for more cascade emission. Furthermore, in the \Fermi-LAT energy range, the leptonic only fit slightly undershoots the highest energy data points for large values of $\delta$, thus a larger cascade luminosity in the \Fermi-LAT energy range reduces the residuals of the fit overall. The radius of the emitting region mildly decreases with the Doppler factor in the EC-DT scan as can be seen in the left panel of figure~\ref{fig:DT_heatmaps}, up to $\delta \sim 30$ and then again for $\delta \gtrsim 40$. Since $L_{p,\rm casc}$, increases almost linearly with $\delta$ and $L_{\gamma}$ is constant, we expect $\xi_{\rm cr, max} \sim 1 / r_b$. This is what we observe in the left and middle panels of figure~\ref{fig:DT_heatmaps}, where scenarios with smaller $r_b$ result in larger $\xi_{\rm cr,max}$ and vice versa. The observed neutrino luminosity follows qualitatively the same trend as $L_{p,\rm casc}$. This is why the number of expected neutrinos in the right panel of figure~\ref{fig:DT_heatmaps} increases with $\delta$. In the above qualitative discussion the exact process (and thus cross section) and the energy dependence of the cross section were neglected. 

The range spanned by $\xi_{\rm cr,max}$ which is $\xi_{\rm cr,max}\leq 20$ is such that the proton luminosity is well below $L_{\rm Edd}$, with $\mathcal{L}_p/L_{\rm Edd} \sim 0.01 - 0.1$ in the $1\sigma$ region around the best-fit $\chi^2$ as summarised in table~\ref{tab:dt_nu_rate}. Interestingly, such values of the baryon loading factor are needed in order for the proton energy budget of blazars to match the local inferred UHECR energy production rate (for example~\cite{2014PhRvD..90b3007M} obtained $\xi_{\rm cr} \sim 3$ assuming $s_p = 2$ considering the entire blazar population). The current upper limit on $\xi_{\rm cr}$ from IceCube is $\xi_{\rm cr} < 5$ for $s_p = 2$~\cite{Aartsen:2016ngq} based on the model of~\cite{2014PhRvD..90b3007M}. However, in that calculation powerful FSRQs produce neutrinos inside the BLR. Relaxing this assumption translates to a less strong limit on $\xi_{\rm cr}$, thus moving the expected limit closer to the values we find. 

We also checked whether the model respects the limit on the parameter $Y_{\nu\gamma}$, introduced in~\cite{Petropoulou:2015upa} for high-spectral-peak blazars, which is defined as the ratio of the 100 MeV - 100 GeV \gRay luminosity of the source $L_{\gamma}^{\rm 0.1 - 100~GeV}$ to the all-flavour neutrino luminosity of the source $L_{\nu}$, $Y_{\nu\gamma} = L_{\gamma}^{\rm 0.1 - 100~GeV}/L_{\nu}$.  The parameter $Y_{\nu\gamma}$ is limited by IceCube observations to values below $Y_{\nu \gamma} \leq 0.13$~\cite{Aartsen:2016ngq} based on the absence of neutrinos in the IceCube extremelly-high energy analysis from the entire high-spectral-peak blazar population. In our model, all scenarios are well below the IceCube limit with $Y_{\nu \gamma} \sim 0.01 - 0.08$. Scenarios with $\delta \sim 50$ which have the largest neutrino luminosity approach $Y_{\nu \gamma} \sim 0.08$. Thus they will be the first to be constrained by IceCube with future observations if the limit on $Y_{\nu\gamma}$ gets stronger. 

The range of radii spanned by the EC-DT model in our scanning range translates to expected variability timescales of order~$\sim 3-4$~days. Thus all the EC-DT models considered here are broadly consistent with the observed variability of \PKS. There is no fundamental reason why $r_b$ cannot be even smaller than what we find here, i.e. smaller than 3~days, except that in our simulations the dissipation radius was fixed to $r_{\rm diss} \sim \delta r_b$ to avoid having degenerate free parameters. 

In principle, the peak temperature of the dust torus could be larger than what we assumed in the preceding analysis, up to $\sim 1000$~K, e.g.~\cite{2011ApJ...732..116M,2007ApJ...660..117C}. If this is the case in \PKS, then the neutrino spectrum would shift to slightly lower energies, in the direction of the peak sensitivity of IceCube, possibly allowing for enhanced neutrino production with respect to the above-quoted results. For 
a simple estimate, we shifted the neutrino energy spectrum to lower energy by a factor of $1000~\mathrm{K}/370$~K~$\sim 2.7$. In this case, the expected neutrino counts increase by a factor of two, but do not otherwise affect our conclusions.

\subsection{Neutrino energy}

Multiplying the expected neutrino flux differential in energy, $\phi_{\nu_{\mu}}$ in eq.~(\ref{eq:Nnu}), with the average GFU effective area in the declination range $\delta = [-5, +30]$~deg we find that the most probable neutrino energy in the EC-DT scenarios studied is $\sim 20$~PeV (in the observer's frame). This is almost two orders of magnitude larger than the most probable energy calculated within IceCube for \icnu. However, the most-probable energy quoted by IceCube was calculated assuming an $\vareps_{\nu}^{-2.19}$ neutrino spectrum, whereas our model predicts a fast-rising neutrino spectrum. 

In order to determine how likely it is that a muon neutrino from the EC-DT model gives rise to a muon that is considered to most probably arise from a 300-TeV neutrino in IceCube assuming an $\vareps_{\mu}^{-2.19}$ neutrino spectrum we must consider the energy losses of muons produced in charged-current interactions outside the IceCube detector. The approach for an ideal detector has been described in Chapter 8 of ~\cite{Gaisser:2016uoy}. 

Muons produced in charged-current interactions outside the detector have initial energies $\vareps_{\mu,0} \approx (1 - y_{\rm CC})\vareps_{\nu_{\mu}}$, where $y_{\rm CC} \sim 0.3$ is the average inelasticity of charged current interactions near 300~TeV which varies weakly with energy~\cite{1996APh.....5...81G}. Thereafer, muons lose energy while traversing rock of length $X$, following~\cite{Lipari:1991ut,Dutta:2000hh}, $\mathrm{d}\vareps_{\mu}/\rm{d} X = -(\alpha + \beta \vareps_{\mu})$, 
with~$\alpha = 2\times 10^{-3}~{\rm GeV~cm^{2}~g^{-1}}$ and $\beta = 5 \times 10^{-6} {\rm cm^2 g^{-1}}$ near 1 PeV. At such high energies the radiative losses ($\beta \vareps_{\mu}$ term) dominate. 

Neutrinos from the direction of \icnu ($z \sim 100.5^{\rm \circ}$) cross a chord of rock with length $ l(z) \sim 2 R_{\oplus} \sin(z - 90^{\circ}) \sim 2322$~km, where $R_{\oplus} \sim 6371$~km is the radius of the Earth. The optical depth for neutrinos with energy $\vareps_{\nu_{\mu}}$ and zenith angle $z$ is $\tau(\vareps_{\nu_{\mu}},z) \sim \sigma_{\nu}(\vareps_{\nu}) \rho l(z)$, where $\sigma_{\nu}$ is the cross section for neutrino charged current interactions, for which we use the parametrisation of~\cite{2018PhRvD..97b3021A} and $\rho$ the density of rock for which we assumed $\rho \sim 2.65{\rm g~cm}^{-3}$. For neutrinos with energy less that 100 PeV, from the direction of \icnu, $\tau(100~{\rm PeV},100.5^{\circ}) \ll 1$. 
Thus, the neutrino interaction probability is approximately constant as a function of length traversed and the distribution of final energies of muons originating from monoenergetic neutrinos interacting over a large distance
through rock is, $\rm{d}N_{\mu}/{\rm d}\vareps_{\mu} \propto \vareps_{\mu}^{-1}$ in the energy range $\vareps_{\mu,0}$ down to $\sim 1$~TeV (see e.g.~\cite{Neronov:2016ksj,Ribordy:2009jj}).

We are, firstly, interested in determining the average muon energy for which the most probable energy of the neutrino is 300 TeV given an $\vareps_{\nu_{\mu}}^{-2.19}$ spectrum.
Considering the steeply falling spectrum, the neutrinos that contribute the most to a particular muon energy $\vareps_{\mu}$ are those with $\vareps_{\nu_{\mu}} = \vareps_{\mu}/y_{\rm CC}$, i.e. the lowest energy neutrinos capable of producing such high energy muons. 
In the present case, this means that the energy of the muon produced by \icnu must have been~$\vareps_{\mu,\rm ref}\sim 210$~TeV. 
We can compare our estimate to the muon-energy estimate of IceCube for \icnuold for which there is more public information (see the supplementary material of~\cite{IceCube:2018dnn}), as the declinations of the two neutrinos are similar ($6^{\circ}$ and $10^{\circ}$ for \icnuold and \icnu respectively) and the reconstructed neutrino energies are also very similar. 
In the case of \icnuold the most probable neutrino energy quoted for an $\vareps_{\nu}^{-2.13}$ ($\vareps_{\nu}^{-2.0}$) spectrum was 290 TeV (311 TeV)
with $90\%$ CL upper limit 4.3 PeV (7.5 PeV). The muon energy proxy was 170 TeV (see top panel of figure S2). Considering that the energy resolution for muons is 0.3 in the log of the energy in this energy range~\cite{IceCube:2013dkx}, our estimate of 210 TeV appears consistent with that of IceCube for \icnuold within the quoted uncertainties. 

We are further interested to know how often a narrow neutrino spectrum with peak at 20 PeV gives rise to a muon with energy $\vareps_{\mu,\rm ref} \leq 210$~TeV. Considering the $\vareps_{\mu}^{-1}$ dependence of the number of muons for monoenergetic neutrino injection, we expect the fraction of muons with energy $\vareps_{\mu,\rm ref} \leq 210$~TeV, $f_{\leq \vareps_{\mu,\rm ref}}$ to be, 
\ba
f_{\leq \vareps_{\mu,\rm ref}} = \frac{\int_{\vareps_{\mu, \rm min}}^{\vareps_{\mu,\rm ref}}{\rm d}N_{\mu}/{\rm d}\vareps_{\mu}~{\rm d}\vareps_{\mu}}{\int_{\vareps_{\mu, \rm min}}^{\vareps_{\mu,\rm max}}{\rm d}N_{\mu}/{\rm d}\vareps_{\mu}~{\rm d}\vareps_{\mu}}
= \frac{\int_{\vareps_{\mu, \rm min}}^{\vareps_{\mu,\rm ref}}{\vareps_{\mu}^{-1}~ {\rm d}\vareps_{\mu}}}{\int_{\vareps_{\mu, \rm min}}^{\vareps_{\mu,\rm max}}{\vareps_{\mu}^{-1}~ {\rm d}\vareps_{\mu}}} \\ \nonumber 
= \frac{\ln(\vareps_{\mu,\rm ref}/\vareps_{\mu,\rm min})}{\ln(\vareps_{\mu,\rm max}/\vareps_{\mu,\rm min})}.
\ea
\noindent Considering monoenergetic injection of 20 PeV neutrinos gives $\vareps_{\mu,\rm max} = 20~\rm PeV(1-y_{\rm CC})$. The minimum muon energy of interest is $\vareps_{\mu,\rm min} = 80(1-y_{\rm CC})~$TeV, where $\sim 80$~TeV is the minimum alert neutrino energy in the Gold Channel transmitted thus far. We find that in the idealised detector considered, $f_{\leq \vareps_{\mu,\rm ref}}\sim 0.25$, in other words $\sim 25\%$ of neutrinos from the EC-DT model would give rise to muons with energy $\leq 210$~TeV. We thus conclude that \icnu is consistent with arising from the EC-DT predicted neutrino model in terms of energy, though additional information from IceCube would allow for a more precise estimate. 

\section{Discussion}
\label{sec:discussion} 

In this work, motivated by the observation of the high-energy neutrino \icnu in the direction of the FSRQ \PKS, and by the fact that this blazar, despite its large redshift, is among the brightest known in terms of their \gRay flux, we performed modelling of the multi-wavelength and neutrino emission from the source, to assess the theoretical grounds for the observed association. 

We considered the ensemble of available astronomical observations of the source, which include \gRays, X-rays, UV, optical, infrared, and radio fluxes, to constrain the theoretical models studied. We considered a comprehensive range of models for the origin of the bulk of the multiwavelength emission from the jet of \PKS, which vary in terms of the distance of the emitting region from the base of the jet, and thus in terms of the available photon fields which act as targets for neutrino production. Motivated by the fact that interferometric radio observations and multi-wavelength (radio-GeV-optical) cross-correlations locate the emitting region of \PKS beyond the broad-line region we investigated scenarios in which the emitting region of \PKS is located beyond the BLR but inside the dust torus (EC-DT model) and beyond the dust torus (SSC model). 

We found that out of the two families of models studied, the long-term SED of \PKS is best described by an emitting region outside the BLR whose high-energy peak is predominantly powered by inverse Compton emission of electrons interacting with the infrared photons emitted by the dust torus. This result is in very good agreement with the ensemble of independent constraints on the location of the emitting region of \PKS from radio and optical observations~\cite{Karamanavis_a,Karamanavis_b,Shao:2020tth} and from earlier searches for the signature of BLR absorption in the \gRay spectra FSRQs which returned a null result for \PKS~\cite{Costamante:2018anp}, as well as with the general result that the majority of FSRQs seem to have emitting regions beyond the BLR~\cite{Costamante:2018anp, Meyer:2019kpy,Harvey:2020wd,acharyya2020locating}. Interestingly, our SED-fitting scan results in a best-fit scenario where the magnetic field strength is $B \sim 0.34-0.38$~G, consistent with independent estimate of~\cite{Shao:2020tth} from optical and radio observations of \PKS. 

The infrared data obtained by the~\emph{AllWISE} data release as part of the \emph{WISE} mission~\cite{2010AJ....140.1868W}, constrain the synchrotron emission of the source and disfavour strongly self-absorbed synchrotron models, such as the leptohadronic model studied for the quiescent spectrum of \PKS in~\cite{Rodrigues:2020fbu}. Our work is the first on the topic of the long-term neutrino emission of \PKS to consider the constraints imposed by the WISE data. Though the WISE data are not strictly simultaneous with the UVOT data which describe the rest of the synchrotron peak of the SED, we demonstrated that they are fully consistent with archival SDSS and 2MASS data and thus that they appear to describe fairly the quiet state of the source. 

An appealing feature of the EC-DT model which we find to be the most compatible with the sum of observational constraints,
is that in the scenario which is most optimistic in terms of neutrino production, the baryon loading is of order 10, and the proton luminosity is well below the Eddington luminosity of the SMBH of \PKS. Such baryon loading is what is expected if blazars as a class power the observed UHECR flux. Furthermore, such value of $\xi_{\rm cr,max}$ is broadly consistent with the upper limit from IceCube as discussed in section~\ref{subsec:DT_result}. This is in contrast to the majority of previous multiwavelength and neutrino emission models in the context of observed neutrino-blazar associations, which required that these blazars produce protons with often super-Eddington luminosity, to get close to detectable neutrino fluxes, see for example the modelling of the 2017 neutrino flare of \TXS~\cite{Ahnen:2018mvi,Cerruti:2018tmc,Gao:2018mnu,Keivani:2018rnh}, as well as the bottom panel of figure 15 of~\cite{Petropoulou:2020pqh}. 

The maximum number of neutrinos that can be expected based on the long-term SED in the EC-DT model is $\mathcal{N}_{\nu_{\mu}+\bar{\nu}_{\mu}}(> 100 ~\rm TeV) \lesssim 0.1/~\rm 10~years$, implying a $\sim 10\%$ probability of detecting one neutrino with the IceCube GFU alert analysis if we assume constant neutrino luminosity. If on the other hand we assume that the neutrino luminosity scales with the \gRay flux of the source then the expected number of muon neutrinos in the same energy range, in ten years is $\sim 1$, and $10\%$ of those neutrinos are expected during \gRay quiet periods such as at the time of arrival of \PKS. The proton-synchrotron model of~\cite{Rodrigues:2020fbu} predicts a similar neutrino flux from \PKS, although the required proton luminosity in that model is significantly higher than in the EC-DT scenario.

Considering the luminosity of this source, and the powerful photon fields that the source is
known to possess (accretion disk, BLR), which could act as target fields for
photopion interactions, it should be possible for it to produce even larger neutrino luminosity and to be in a regime interesting for detection of $\sim$ few neutrinos with IceCube even with a modest baryon loading (see for
example~\cite{Atoyan:2001ey,Dermer:2014vaa,2014PhRvD..90b3007M,Rodrigues:2017fmu}), if the dissipation happens close to the base of the jet. An important advantage of neutrino production in photomeson interactions with BLR photons is that the energy of the BLR photons is such that neutrino production is expected at energy $\vareps_{\nu} \sim 300~\mathrm{TeV}(10~\mathrm{eV}/\vareps_{\rm BLR})/(1+z)$, where $\vareps_{\rm BLR}\sim 10~$eV is the typical rest-frame energy of BLR photons, which is very favourable for detection in IceCube. Such a scenario was investigated by~\cite{Rodrigues:2020fbu}. In the BLR model, \PKS could produce such neutrino flux as to trigger the detection of $N_{\nu_{\mu}} \sim  2^{+8}_{-2}$ and $N_{\nu_{\mu}} \sim  11\pm 6$ muon neutrinos in the IceCube Point Source selection during quiet periods, and during the 2008 and subsequent hard-X-ray flares up until 2018 respectively. However, the fact that the eight-year (2009-2017) IceCube Point Source analysis~\cite{Aartsen:2018ywr} and IC40 Point source analysis~\cite{Abbasi_2011} both find no neutrinos from the direction of \PKS between 2008-2019, rules out neutrino emission at such a high rate in this energy range. The most optimistic SSC models we have investigated in section~\ref{sec:results} are similarly ruled out by the absence of archival neutrino emission. 

The absence of a strong neutrino signal from FSRQs so far from stacking~\cite{Aartsen:2016lir,Huber:2019lrm}, clustering~\cite{Murase:2016gly,Yuan:2019ucv} and diffuse limits at higher neutrino
energies~\cite{Aartsen:2018vtx,Murase:2018iyl} in IceCube, is consistent with the results of~\cite{Costamante:2018anp}
which suggests that in the majority of FSRQs the $\gamma$-ray emitting region is beyond the BLR and thus a strong neutrino signal at sub-PeV energies is not expected. Their results, as well as our detailed
modelling of \PKS as a case study thus illustrate that the location of the FSRQ emitting region beyond the BLR on average may be a
crucial reason why FSRQs are not yet discovered as IceCube point sources, despite their otherwise very powerful 
jets, assuming that the neutrino and $\gamma$-ray emitting regions are cospatial (see also~\cite{2014PhRvD..90b3007M,Rodrigues:2017fmu,Righi:2020ufi}). 

At the time of arrival of \icnu, \PKS was in a quiet state in the optical to \gRay energy range, but radio observations reveal that it was experiencing an all-time high at 15~GHz as measured with the OVRO 40m telescope~\cite{ovro_atel}. Intriguingly, a similar pattern was observed in the OVRO data of \TXS at the time of arrival of \icnuold~\cite{ovro_atel}. Even though the similarity of these observations is at first sight intriguing, there are also important differences among the two cases, for example, \TXS was experiencing a strong \gRay flare in 2017, whereas \PKS was in a long-term quiet state. 

The analysis of~\cite{Karamanavis_b} locates the 15 GHz emission of \PKS at $\sim 6$~pc from the base of the jet and the \gRay emitting region at $1.9\pm1.1$~pc. Neutrinos and \gRays are naturally cospatially produced in photopion interactions, we have thus focussed on this possibility in this work. Unlike 2008 when the strong 15 GHz flare of \PKS was preceded by a strong \gRay flare, no \gRay flare was observed in 2019 (see e.g. figure 8 of~\cite{2021MNRAS.503.3145B}). Thus, there is no obvious way to connect the 2019 OVRO observations to the expected neutrino emission in our model. 

The possible relevance of the 15 GHz high state of \PKS and \TXS at the time of arrival of \icnu and \icnuold respectively has been addressed by~\cite{Kun:2020njy} who proposed a possible mechanism that explains the observation of high energy neutrinos with a radio high state and a brief \gRay dimming of the sources. However, thus far it has not been estimated quantitatively whether in this or similar scenario \TXS or \PKS would be able to produce sufficient neutrino flux to explain the observation of \icnuold and \icnu respectively. Qualitatively, if the long term radio outburst which started for \PKS in 2014 signifies a large outflow from the source as was suggested by~\cite{Kun:2020njy}, it may be coupled to a larger-than-average proton loading which is favourable for neutrino production in our model and in general. In our work, the proton content of the jet is a semi-free parameter which is bounded by the electromagnetic cascade emission expected from the interactions of the protons.

If conditions such as those studied in this work are ubiquitous among FSRQs, then we expect that the stacked
neutrino signal from these sources which peaks at neutrino energy beyond $\sim \mathcal{O}(10~\rm PeV)$ is within
reach of IceCube with increased exposure and that FSRQs will be scrutinised as neutrino sources with neutrino detectors optimised in the sub-EeV energy range including the next generation radio neutrino observatory~\cite{Aguilar:2019jay}, and eventually 
IceCubeGen2-radio~\cite{IceCube-Gen2:2020qha}, and GRAND~\cite{GRAND:2018iaj}, and optical facilities optimised in the sub-EeV energy range such as the proposed TRINITY~\cite{Otte:2019aaf}, top-of-the-mountain fluorescence-telescope system~\cite{Neronov:2019htv}, and POEMMA~\cite{POEMMA:2020ykm}. 

\acknowledgments
We acknowledge useful discussions with Theo Glauch, Matthias Huber, Hans Niederhausen, Xavier Rodrigues and Michael Unger. We thank Anna Franckowiak and Simone Garrappa for providing us with the \Fermi-LAT lightcurve of \PKS. This publication makes use of data products from the Wide-field Infrared Survey Explorer, which is a joint project of the University of California, Los Angeles, and the Jet Propulsion Laboratory/California Institute of Technology, funded by the National Aeronautics and Space Administration. SB acknowledges financial support by the European Research Council for the ERC Starting grant MessMapp, under contract no. 949555. MP acknowledges support from the MERAC Fondation through the project THRILL, and by the Deutsche Forschungsgemeinschaft (DFG, German Research Foundation) through Grant Sonderforschungsbereich (Collaborative Research Center) SFB1258 ``Neutrinos and Dark Matter in Astro- and Particle Physics'' (NDM) as a Mercator Fellow. MS acknowledges support by the National Science Foundation through grants PHY-1914579 and PHY-1913607. 

\bibliography{pks1506}

\appendix
\section{Calculation of cascade emission}
\label{sec:cascade} 

High-energy protons, if present in copious amounts in the
jet of \PKS, will lead to the production not only of neutrinos but also of 
hadronic \gRays and high-energy electrons, primarily through photomeson ($p\gamma$) interactions and Bethe-Heitler pair production. The high-energy electrons radiate synchrotron and IC photons, adding to the observed emission of the blazar, while additional electron/positron pairs are produced by the 
$\gamma$-rays interacting with the radiation fields in the source environment. 
In this work, we have taken the approach of fitting the SED of \PKS with a leptonic model, which amounts to the assumption that the majority of observed emission is of leptonic origin, and introducing an additional proton component whose maximum luminosity is constrained by the multiwavelength SED of \PKS. The sum of the leptonic emission and the additional radiation from proton interactions cannot exceed the total observed luminosity of \PKS at any energy. In this section, we describe how the secondary radiation emitted by proton interactions has been calculated. 

Of the energy lost by protons in $p\gamma$ interactions, 
$3/8$ths go to the production of neutrinos, following eq.~(\ref{eq:nuLuminosity}). The remaining $5/8$ths of the energy lost result in 
the production of electrons and $\gamma$-rays. Gamma-rays are produced by the decay of neutral pions, with luminosity
\be
\vareps_{\gamma} L_{\vareps_{\gamma}} = \frac{1}{2} f_{p\gamma}(\vareps_{p}) \vareps_{p} L_{\vareps_{p}} = \frac{4}{3} \vareps_{\nu} L_{\vareps_{\nu}}|_{\vareps_{\gamma}=2 \vareps_{\nu}},
\ee
and electrons are produced by charged pion decay with luminosity,  $\vareps_{e} L_{\vareps_{e}} = 1/3\,\vareps_{\nu} L_{\vareps_{\nu}}|_{\vareps_e = \vareps_{\nu}}$. 

The \gRays may experience 
further interactions inside the source if the optical depth for electron-positron 
pair production is appreciable. Otherwise they escape the source. 
The \gRays that do escape the source with energy $\gtrsim 100/(1+z)$~GeV are attenuated by the EBL. 
The optical depth to photon-photon pair production is given by,
\be
\tau_{\gamma \gamma}(\vareps^{\prime}_{\gamma}) \approx r_{b} \int^{\vareps^{\prime}_{t,min}}_{\vareps^{\prime}_{t,max}} \sigma_{\gamma \gamma}(\vareps^{\prime}_t,\vareps^{\prime}_{\gamma}) n^{\prime}_{\vareps^{\prime}_t}(\vareps^{\prime}_{\gamma})
\ee
\noindent where $n^{\prime}_{\vareps_{t}}$ is the number density of target photons with energy $\varepsilon^{\prime}_t$ and we assume a homogeneous photon field inside the source. The cross section for the process, $\sigma_{\gamma \gamma}(\vareps^{\prime}_t,\vareps^{\prime}_{\gamma})$, is assumed to follow the analytical form given by eq. (3.23) of~\cite{2004vhec.book.....A}. Both the internal and the external photon fields are considered in the calculation. The attenuated \gRays produce electron-positron pairs, which are assumed to each have energy 
$\vareps^{\prime}_e \approx \vareps^{\prime}_{\gamma}$, in other words, one of the produced leptons takes almost all the photon energy, when $\vareps^{\prime}_{t} \vareps^{\prime}_{\gamma}/m_e^2 c^4 \gtrsim 3$. The condition is satisfied by the very highly energetic photons produced in neutral pion decays. The electrons that are subsequently produced have luminosity,
\be
\vareps^{\prime}_e L^{\prime}_{\vareps_e} \approx \vareps^{\prime}_{\gamma} L^{\prime}_{\vareps^{\prime}_{\gamma}}|_{\vareps^{\prime}_e \approx \vareps^{\prime}_{\gamma}} [1 - \exp{(-\tau_{\gamma\gamma}(\vareps^{\prime}_{\gamma})}].
\label{eq:pairs_from_gg} 
\ee
At lower centre-of-mass energies, $\vareps^{\prime}_{t} \vareps^{\prime}_{\gamma}/m_e^2 c^4 < 3$, each of the electrons are assumed to be produced with energy $\vareps^{\prime}_e \approx \vareps^{\prime}_{\gamma}/2$ and eq.~(\ref{eq:pairs_from_gg}) is modified accordingly. 

Below the threshold for $p\gamma$ interactions, protons lose energy through the production of electron-positron pairs (Bethe-Heitler, hereafter BH, process). The timescale for BH pair-production is calculated in the same way as the $p\gamma$ timescale, which is given by eq.~(\ref{eq:pgammaRate}), except for a different cross section and inelasticity. 

For the cross section of the BH process, we use the numerical fit given by eq. (9) of~\cite{1983MNRAS.204.1269S}. For the inelasticity of the process we use the approximate expression $\kappa_{\rm BH} \approx m_e / m_p$, e.g.~\cite{Dermer:2010hy}. Thus, each produced electron is assumed to have energy, $\vareps^{\prime}_{\rm BH} = \kappa_{\rm BH} \vareps_p^{\prime}$. We have checked the validity of our approximate treatment by comparing the results of those obtained with numerical codes and find that the total number of injected electrons (and total injected energy) agree with full numerical results within better than a factor of two~\cite{Cerruti:2021hah}.  

To obtain the steady-state electron spectrum inside the emitting blob resulting from the electrons produced by BH, photomeson and photon-photon pair production processes, we use the steady-state transport equation, 
\be
\frac{\mathrm{d}}{\mathrm{d}\vareps_e^{\prime}} \left[\frac{\mathrm{d}\vareps_e^{\prime}}{\mathrm{d}t^{\prime}} N^{\prime}(\vareps_e^{\prime})\right] = -\frac{N^{\prime}(\vareps_e^{\prime})}{t^{\prime}_{\rm esc}}+\dot{N}^{\prime} (\vareps_e^{\prime}),
\label{eq:continuity}
\ee
where $N^{\prime}(\vareps^{\prime}_e)$ is the electron number spectrum, $\dot{N}^{\prime} (\vareps^{\prime}_e)$ is the source term, $t^{\prime}_{\rm esc}$ is the escape time which we assume to be equal to the blob crossing time, $t^{\prime}_{\rm cross}$ and $\mathrm{d}\vareps^{\prime}_e/\mathrm{dt^{\prime}} \approx \mathrm{d}\vareps^{\prime}_e/\mathrm{dt^{\prime}}_{\rm IC} + \mathrm{d}\vareps^{\prime}_e/\mathrm{dt^{\prime}}_{\rm syn}$ the energy loss rate assumed to be dominated by synchrotron and inverse Compton energy losses. This equation has the analytical solution~\cite{Dermer:2010hy}, 
\be
N^{\prime}(\vareps_e^{\prime}) = \left|\frac{\mathrm{d}\vareps^{\prime}_e}{\mathrm{d}t}\right|^{-1} \int^{\infty}_{\vareps^{\prime}_e}  \mathrm{d} \vareps_e^{\prime} \dot{N}^{\prime} (\vareps_e^{\prime}) \exp \left( - \frac{1}{t^{\prime}_{\rm cross}} \int_{\vareps^{\prime}_e}^{\vareps_e^{\prime \prime}} \frac{\mathrm{d} {\vareps_e}^{\prime \prime}}{\left| \mathrm{d}\vareps^{\prime}_e/\mathrm{d}t(\vareps_e^{\prime \prime})\right|}\right)
\label{eq:soln}
\ee
We numerically integrate eq.~(\ref{eq:soln}) to obtain the steady-state electron spectrum resulting from to photo-hadronic processes. 

For the radiation processes of these high-energy electrons, we consider synchrotron and inverse Compton emission. We calculate the energy lost by these electrons to synchrotron radiation as, 
\be
\frac{{\rm d}\vareps_e^{\prime}}{{\rm d}t_{\rm syn}^{\prime}} = - \frac{4}{3}\sigma_{\rm T  }c \gamma^{\prime 2}_e U_B^{\prime}
\ee
\noindent where $U_B^{\prime} = B^{\prime 2}/8\pi$ is the energy density of the magnetic field, and via inverse Compton emission with rate ${\rm d}\vareps^{\prime}_e/{\rm d}t_{\rm IC}$ which we calculate using eq. (2.56) of~\cite{1970RvMP...42..237B}. The synchrotron energy radiated per unit photon energy is calculated following eq. (9) of~\cite{1988ApJ...334L...5G}. The spectrum of photons radiated via the IC process is calculated using eq. (2.48) and (2.61) of~\cite{1970RvMP...42..237B}.  
The electrons and positrons undergo additional IC and synchrotron cooling. To calculate this additional ``cascade'' emission, we solve eq.~(\ref{eq:soln}) iteratively, adding the extra cascade radiation to the electron source term in each step until the integral of the electron distribution, $N^{\prime}(\vareps^{\prime}_e)$ converges to a value which changes by less than $10\%$ in subsequent iterations. We have checked that our results do not change if we consider instead the integral of $N^{\prime}(\vareps^{\prime}_e)\vareps^{\prime}_e$ for the stopping condition. 

For each parameter set explored in section~\ref{sec:results}, we check that this, total cascade emission as described in this section, boosted to the observer frame, does not exceed the observed emission of \PKS at any wavelength, i.e. it doesn't degrade the \chisq of the fit by more than 4 which is a $2\sigma$ increase, otherwise, we lower the proton luminosity of the source. 

\end{document}